\documentclass[10pt,twocolumn,aps,pra,showpacs,superscriptaddress,floatfix,nofootinbib]{revtex4-1}

\usepackage[T1]{fontenc}
\usepackage{graphicx}
\usepackage{amssymb}
\usepackage{amsmath}
\usepackage{color}
\usepackage{psfrag}
\usepackage{epsfig}
\usepackage{bbm}
\usepackage{bm}
\usepackage{hyperref}
\usepackage[normalem]{ulem}
\usepackage{amsthm}
\usepackage{epstopdf}

\definecolor{nred}{rgb}{0.9,0.1,0.1}
\definecolor{nblack}{rgb}{0,0,0}
\definecolor{nblue}{rgb}{0.2,0.2,0.8}
\definecolor{ngreen}{rgb}{0.2,0.6,0.2}

\newcommand{\blu}{\color{nblack}}

\usepackage{etoolbox}
\usepackage{bbold} 

\newcommand{\ket}[1]{| #1 \rangle}
\newcommand{\bra}[1]{\langle #1 |}
\newcommand{\proj}[1]{\ket{#1}\!\bra{#1}}
\newcommand{\braket}[1]{\langle  #1 \rangle}

\newcommand{\beq}{\begin{eqnarray}}
\newcommand{\eeq}{\end{eqnarray}}

\newcommand{\id}{\openone}
\newcommand{\SCHSH}{\mathcal{S}_{\mbox{\tiny CHSH}}}

\newcommand{\I}{\mathcal{I}}

\newcommand{\Q}{\mathcal{Q}}

\newcommand{\ts}{\tilde{\sigma}}
\newcommand{\tGl}{\tilde{G}_\lambda}

\newcommand{\tA}{\text{A}}
\newcommand{\tB}{\text{B}}
\newcommand{\tC}{\text{C}}
\newcommand{\ttC}{\text{\tiny C}}

\newcommand{\rab}{{\rho_{\mbox{\tiny AB}}}}
\newcommand{\rc}{{\rho_\ttC}}
\newcommand{\rabc}{{\rho_{\mbox{\tiny ABC}}}}
\newcommand{\rId}{\rho_{\text{\tiny I},d}}
\newcommand{\rIdv}{\rId(v_d)}

\newcommand{\tab}{{\tau_{\mbox{\tiny AB}}}}
\newcommand{\oab}{{\omega_{\mbox{\tiny AB}}}}

\newcommand{\rabxy}{{\rho_{ab|xy}^{\mbox{\tiny C}}}}
\newcommand{\vecP}{\mathbf{P}}
\newcommand{\Pobs}{\mathbf{P}_\text{obs}}
\newcommand{\Pobsabcxyz}{{P_{\mbox{\tiny obs}}(abc|xyz)}}

\newcommand{\ER}{{\rm ER}}
\newcommand{\NR}{{\rm NR}}
\newcommand{\IR}{{\rm IR}}
\newcommand{\NRcPabxy}{\NR^\text{c}(\vecP)}

\newcommand{\NRPabxy}{\NR(\{P(a,b|x,y)\})}
\newcommand{\NRPobs}{\NR(\mathbf{P}_{\text{obs}})}
\newcommand{\NRc}{\NR^\text{c}}

\newcommand{\chil}{\chi^{(\ell)}}
\newcommand{\chilbb}{\chi_{\mbox{\tiny DI}}^{(\ell)}}
\newcommand{\chiDI}{\chi_{\mbox{\tiny DI}}^{(\ell)}[\rho_{a|x}]}
\newcommand{\chiDIfirst}{\chi_{\mbox{\tiny DI}}^{(1)}[\rho_{a|x}]}
\newcommand{\chiDIabc}{\chi_{\mbox{\tiny DI}}^{(\ell)}[\rho_{ab|xy}^{\mbox{\tiny C}}]}
\newcommand{\chiDIabcfirst}{\chi_{\mbox{\tiny DI}}^{(1)}[\rho_{ab|xy}^{\mbox{\tiny C}}]}

\newcommand{\SRDIPobs}{{\rm SR}_{\mbox{\tiny DI},\ell}^{\mbox{\tiny A$\rightarrow$B}}(\mathbf{P}_{\text{obs}})}

\newcommand{\SWDIPobs}{{\rm SW}_{\mbox{\tiny DI},\ell}^{\mbox{\tiny A$\rightarrow$B}}(\mathbf{P}_{\text{obs}})}
\newcommand{\SRDI}{{\rm SR}_{\mbox{\tiny DI},\ell}}

\newcommand{\SRDIcPobs}{{\rm SR}_{\mbox{\tiny DI},\ell}^{\mbox{\tiny c,A$\rightarrow$B}}(\mathbf{P}_{\text{obs}})}
\newcommand{\SWDIcPobs}{{\rm SW}_{\mbox{\tiny DI},\ell}^{\mbox{\tiny c,A$\rightarrow$B}}(\mathbf{P}_{\text{obs}})}
\newcommand{\SRDIc}{{\rm SR}_{\mbox{\tiny DI},\ell}^{\mbox{\tiny c,A$\rightarrow$B}}}

\newcommand{\SRc}{{\rm SR}^{\text{c}}}
\newcommand{\SRrax}{{\rm SR}(\{\rho_{a|x}\})}
\newcommand{\SRcrax}{{\rm SR}^{\text{c}}(\{\rho_{a|x}\})}
\newcommand{\IREax}{{\rm IR}(\{E_{a|x}^{\mbox{\tiny A}}\})}
\newcommand{\IREby}{{\rm IR}(\{E_{b|y}^{\mbox{\tiny B}}\})}
\newcommand{\IWEax}{{\rm IW}(\{E_{a|x}^{\mbox{\tiny A}}\})}

\newcommand{\SWrax}{{\rm SW}(\{\rho_{a|x}\})}

\DeclareMathOperator{\tr}{tr}

\theoremstyle{definition}

\begin{document}

\title{Exploring the framework of assemblage moment matrices and its applications in device-independent characterizations}

\author{Shin-Liang Chen}
\email{shin-liang.chen@mpq.mpg.de}
\affiliation{Max-Planck-Institut f{\"u}r Quantenoptik, Hans-Kopfermann-Stra{\ss}e 1, 85748 Garching, Germany}
\affiliation{Department of Physics, National Cheng Kung University, Tainan 701, Taiwan}
\author{Costantino Budroni}
\email{costantino.budroni@oeaw.ac.at}
\affiliation{Institute for Quantum Optics and Quantum Information (IQOQI), Austrian Academy of Sciences, Boltzmanngasse 3 1090 Vienna, Austria}
\author{Yeong-Cherng Liang}
\email{ycliang@mail.ncku.edu.tw}
\affiliation{Department of Physics, National Cheng Kung University, Tainan 701, Taiwan}
\author{Yueh-Nan Chen}
\email{yuehnan@mail.ncku.edu.tw}
\affiliation{Department of Physics, National Cheng Kung University, Tainan 701, Taiwan}
\affiliation{Physics Division, National Center for Theoretical Sciences, Hsinchu 300, Taiwan}

\date{ \today}

\begin{abstract}
In a recent work [Phys. Rev. Lett. \textbf{116}, 240401 (2016)], a framework known by the name of \emph{assemblage moment matrices} (AMMs) has been introduced for the device-independent quantification of quantum steerability and measurement incompatibility. In other words, even with no assumption made on the preparation device nor the measurement devices, one can make use of this framework to certify, directly from the observed data, the aforementioned quantum features. Here, we further explore the framework of AMM and provide improved device-independent bounds on the generalized robustness of entanglement, the incompatibility robustness and the incompatibility weight. We compare the tightness of our device-independent bounds against those obtained from other approaches. Along the way, we also provide an analytic form for the generalized robustness of entanglement for an arbitrary two-qudit isotropic state. When considering a Bell-type experiment in a tri- or more-partite scenario, we further show that the framework of AMM provides a natural way to characterize a superset to the set of quantum correlations, namely, one which also allows post-quantum steering.

\end{abstract}

\maketitle

\section{Introduction}

By using a \emph{Bell-nonlocal}~\cite{Bell64,NLreview} resource, such as an entangled pure quantum state, one can generate correlations between measurement outcomes which do not obey the principle of local causality~\cite{Bell04}, beating our intuitive understanding of nature. To date, convincing experimental demonstrations of Bell-nonlocality (hereafter abbreviated as nonlocality) have been achieved in a number of different physical systems (see, e.g., Refs.~\cite{Hensen15,Shalm15,Giustina15,rosenfeld_event-ready_2017}). 

Operationally, nonlocality  enables one to perform some tasks that are not achievable in classical physics, including quantum cryptography~\cite{Ekert91}, randomness generation~\cite{Colbeck2006,Pironio10}, reduction of communication complexity~\cite{Cleve97} etc. For example, using nonlocal correlations, the task of quantum key distribution~\cite{Ekert91} can be achieved~\cite{Acin07} even when one assumes nothing about the shared quantum resource or the measurement apparatuses. Since then, several quantum information tasks have been proposed within this black-box paradigm (see \cite{Brunner08,Scarani12,Pironio16} and references therein) --- forming a discipline that has come to be known as \emph{device-independent (DI) quantum information}.

Another peculiar feature offered by quantum theory is \emph{steering}~\cite{Schr35_0} --- the fact that one can remotely steer the set of conditional quantum states (called an \emph{assemblage}~\cite{Pusey13}) accessible by a distant party by locally measuring a shared entangled state. This intriguing phenomenon was revisited in 2007 by Wiseman, Jones, and Doherty~\cite{wiseman2007}. In turn, their mathematical formulation forms the basis of a very active field of research (see, e.g., Refs.~\cite{Cavalcanti09,SNC14,Piani15,Gallego15,HLL2016} and references therein) and has given rise to the so-called {\em one-sided DI quantum information}~\cite{Branciard12}.

To exhibit nonlocality or to demonstrate the steerability of a quantum state, it is necessary to employ \emph{incompatible measurements}~\cite{Wolf09}. In particular, among existing formulations of such measurements~\cite{Heinosaari10,Reeb13,Haapasalo15}, {\em any} measurements that are incompatible---in the sense of being \emph{non-jointly-measurable}~\cite{Liang:PRep}---can always be used~\cite{Quint14,Uola14} to demonstrate the steerability of some quantum states. In fact, {\blu the} incompatibility robustness~\cite{Uola15}---a  quantifier for measurement incompatibility---has even been shown to be lower bounded~\cite{SLChen16,Cavalcanti16} by {\blu the} steering robustness~\cite{Piani15} -- a  quantifier for quantum steerability.

In the context of DI quantum information, a \emph{moment matrix}, i.e., a matrix composed of a set of expectation values of observables, is known to play a very important role. In particular, the hierarchy of moment matrices due to  Navascu\'{e}s, Pironio, and Ac\'{i}n (NPA)~\cite{NPA} not only has provided the only known effective characterization (more precisely, approximation) of the quantum set, but also has found applications in DI entanglement detection~\cite{Bancal11,Baccari17}, quantification~\cite{Moroder13,Liang15,SLChen16}, dimension-witnessing~\cite{Brunner08,Navascues14,Navascues15prl}, self-testing~\cite{Yang14,Bancal15} etc. 
Similarly, some other variants~\cite{Pusey13,Kogias15} of the NPA hierarchy have also found applications in the context of one-sided DI quantum information. In Appendix~\ref{Sec_App_MMs}, we summarize in Table \ref{TB_MM_Refs} some of the hierarchy of moment matrices that have been considered in (one-sided) DI quantum information.

Inspired by the moment matrices considered in Refs.~\cite{Moroder13,Pusey13}, a framework known by the name of \emph{assemblage moment matrices} (AMMs) was proposed in Ref.~\cite{SLChen16}. As opposed to previous considerations, a distinctive feature of AMM is that the moment matrices considered {\blu consist} of expectation values only for subnormalized quantum states (specifically, the assemblage induced in a steering experiment). This unique feature makes AMM a very natural framework for the DI quantification of steerability, and consequently the DI quantification of measurement incompatibility as well as the DI quantification of entanglement robustness, and its usefulness in certain quantum information tasks. 

In this paper, we further explore the relevance of AMM for DI characterizations. We begin in Sec.~\ref{Sec_MM} by reviewing the concept of moment matrices considered in DI quantum information. Then, we recall from Ref.~\cite{SLChen16} the framework of AMM in Sec.~\ref{Sec:AMM}. After that, we discuss the applications of AMM in DI quantum information, specifically DI characterizations. In Sec.~\ref{Sec:Conclusion}, we conclude with a summary results and outline some possibilities for future research.

\section{Moment matrices within the device-independent paradigm}
\label{Sec_MM}

Moment matrices, i.e., matrices of expectation values of certain observables, were first discussed in a DI setting by NPA in Ref.~\cite{NPA}. For our purposes, however, it would be more convenient to think about these matrices as the result of some local, complete-positive (CP) maps acting on the underlying density matrix, as discussed in Ref.~\cite{Moroder13}. To this end, consider two local CP maps $\Lambda_\text{A}$ and $\Lambda_\text{B}$ acting, respectively, on  Alice's and Bob's system ($\rho_\text{A}$ and $\rho_\text{B}$):
\begin{subequations}\label{Eq:LocalMapping}
\begin{equation}
\begin{aligned}
	\Lambda_{\text{A}}(\rho_{\text{A}})= \sum_n K_n \rho_{\text{A}} K_n^\dagger, \quad
	\Lambda_{\text{B}}(\rho_{\text{B}}) = \sum_m L_m \rho_{\text{B}} L_m^\dagger,
\end{aligned}
\end{equation}
where the Kraus operators are
\begin{equation}
	K_n = \sum_i |i\rangle_{\bar{\text{A}}\text{A}} \langle n|A_i, \quad
	L_m = \sum_j |j\rangle_{\bar{\text{B}}\text{B}} \langle m|B_j,
\label{Eq_kraus}
\end{equation}
\end{subequations}
while $\{\ket{i}_{\bar{\text{A}}}\}, \{\ket{n}_{\text{A}}\}$ ($\{\ket{j}_{\bar{\text{B}}}\}, \{\ket{m}_{\text{B}}\}$) are, respectively, orthonormal bases for the output Hilbert space $\bar{\text{A}}$ ($\bar{\text{B}}$) and input Hilbert space A (B) of Alice's (Bob's) system. In Eq.~\eqref{Eq_kraus}, $A_i$ and $B_j$ are, respectively, operators acting on Alice's and Bob's input Hilbert space.

Together, when applied to a quantum state $\rab$, these local CP maps give rise to a matrix $\chi$ of expectation values $\braket{A_k^\dagger A_i \otimes B_l^\dagger B_j}_\rab$ 
\begin{equation}
\begin{split}
&\,\chi[\rab,\{A_i\} ,\{B_j\}] \\
&= \Lambda_{\text{A}}\otimes\Lambda_{\text{B}}(\rab)\\
&= \sum_{ijkl}\ket{ij}\!\bra{kl}\tr[\rab A_k^\dagger A_i \otimes B_l^\dagger B_j],
\end{split}
\label{Eq_MM1}
\end{equation}
which is a function of $\rab$, as well as the choice of  $\{A_i\}$ and $\{B_j\}$.

Consider now a bipartite Bell experiment where Alice (Bob) can freely choose to perform any of the $n_x$ ($n_y$) measurements, each giving $n_a$ ($n_b$) possible outcomes. In quantum theory, these measurement are described by positive-operator-valued measures (POVMs). Let $\{E^\tA_{a|x}\}_{x,a}$ and $\{E^\tB_{b|y}\}_{y,b}$ respectively denote the collection of POVM elements (also known as a {\em measurement assemblage}~\cite{Piani15}) associated with Alice's and Bob's measurements, and let $\id$ be the identity operator. Then, if we let $\{A_i\}$ ($\{B_j\}$) be the set of operators obtained by taking all $\ell$-fold products of operators from $\{\id\}\cup\{E^\tA_{a|x}\}_{x,a}$ ($\{\id\}\cup\{E^\tB_{b|y}\}_{y,b}$), the corresponding moment matrix, cf. Eq.~\eqref{Eq_MM1}, is said~\cite{Moroder13} to be a moment matrix of {\em local} level $\ell$ (see also Ref.~\cite{Vallins17}). Note that for all $\ell\ge 1$, one can find in  the corresponding moment matrix $\chil$ expectation values that are (at most) first order in $E^\tA_{a|x}, E^\tB_{b|y}$. From Born's rule, one finds that they correspond to the joint probability of Alice (Bob) observing outcome $a$ ($b$) conditioned on she (he) performing the $x$-th ($y$-th) measurement, i.e., 
\begin{equation}\label{Eq:Quantum}
	P(a,b|x,y)\stackrel{\Q}{=}\tr\left(\rab\,E^\tA_{a|x}\otimes E^\tB_{b|y}\right). 
\end{equation}
Importantly, these quantities can be estimated directly from the experimental data without assuming any knowledge about the POVM elements nor the shared state $\rab$. In addition, all legitimate moment matrices of the form of Eq.~\eqref{Eq_MM1} are easily seen to be positive semidefinite, denoted by $\chi\succeq 0$. Thus, in a DI paradigm when only the correlations $\mathbf{P}_\text{obs}=\{P(a,b|x,y)\}_{a,b,x,y}$ are assumed (or estimated), one can still determine through the positive semidefinite nature of moment matrices if $\mathbf{P}_\text{obs}$ is {\blu not} quantum realizable. 

Let us denote by $\chilbb$ the corresponding moment matrix in this black-box setting. If there is no way to fill in the remaining unknown entries of $\chilbb$ [collectively denoted by $\{u_i\}$] such that $\chilbb\succeq 0$, one would have found a certificate showing that the given $\mathbf{P}_\text{obs}$ is {\em not} quantum realizable [in the sense of Eq.~\eqref{Eq:Quantum}].  From these observations, a hierarchy~\cite{NPA2008,Doherty08,Moroder13} of superset approximations $\tilde{\Q}^{(\ell)}$ to the set of legitimate quantum correlations (denoted by $\Q$) can be obtained by solving a hierarchy of semidefinite programs, each associated with a moment matrix of local level $\ell$. Moreover, the hierarchy $\tilde{\mathcal{Q}}^{(1)}\supsetneq \tilde{\mathcal{Q}}^{(2)}\supsetneq ...\supsetneq \tilde{\mathcal{Q}}$ provably converges to $\Q$, i.e., $\tilde{\mathcal{Q}}^{(\ell\rightarrow\infty)}\rightarrow\mathcal{Q}$ (see also~\cite{NPA2008,Doherty08}). In performing this algorithmic characterization, since any POVM can be realized as a projective measurement (embedded in higher-dimensional Hilbert space~\cite{Neumark}), without loss of generality one can thus set the uncharacterized $\{E_{a|x}\}_a$ and $\{E_{b|y}\}_b$ to be projectors for all $x$ and $y$, such that $E_{a|x}E_{a'|x}=\delta_{a,a'}E_{a|x}$ and $E_{b|y}E_{b'|y}=\delta_{b,b'}E_{b|y}$. In addition, one can further assume that each $u_i$ is a real number; see~\cite{Moroder13} for the detailed reasonings behind these simplifications. In Table~\ref{TB_MM}, we provide a summary of the various elements of $\chilbb$ in relation to the operators whose expectation values are to be evaluated.

\begin{center}
\begin{table}[h!]
\centering
\caption{Elements of the moment matrix $\chilbb$ constructed from Eq.~\eqref{Eq_MM1} with the simplification that all measurements are described by orthogonal projectors.  
}
\begin{tabular}{|c|c|}
\hline
elements & for $A_k^\dagger A_i$ ($B_l^\dagger B_j$) \\ \hline \hline
0 & containing $E_{a|x}^\tA E_{a'|x}^\tA$ with $a\neq a'$ \\
&  (or $E_{b|y}^\tB E_{b'|y}^\tB$ with $b\neq b'$) \\ \hline
$P_\text{obs}(a,b|x,y)$ & being $E_{a|x}^\tA$  (and $E_{b|y}^\tB$) \\ \hline
unknown $u_i\in\mathbb{R}$ & being otherwise \\ \hline
\end{tabular}\label{TB_MM}
\end{table}
\end{center}

\section{Assemblage moment matrices \& quantum steering}
\label{Sec:AMM}


\subsection{Steerability}

In the DI paradigm explained above,  all preparation devices and measurement devices are treated as uncharacterized (black) boxes. In contrast, consider now a situation where the measurements devices of one party, say, Bob, are fully characterized. Then, for every outcome $a$ that Alice obtains when she performs the $x$-th measurement, Bob can in principle perform quantum state tomography to determine the corresponding quantum state $\hat{\rho}_{a|x}$ prepared on his end. 

In quantum theory, if the shared quantum state is $\rab$ and Alice's measurement assemblage is given by $\{E_{a|x}^\tA\}_{a,x}$ (henceforth abbreviated as $\{E_{a|x}^\tA\}$), then $\hat{\rho}_{a|x}$ is simply the normalized version of the conditional state
\begin{equation}\label{Eq_quantum_assemblage}
	\rho_{a|x} = \tr_\text{A}(E_{a|x}^\tA\otimes\mathbb{1}~\rab)\quad \forall\,\, a,x,
\end{equation}
where $\tr_\text{A}(.)$ refers to a partial trace over Alice's Hilbert space. Explicitly, if we denote by $P(a|x)=\tr(\rho_{a|x})$, then $\hat{\rho}_{a|x}={\rho}_{a|x}/P(a|x)$. Following Ref.~\cite{Pusey13}, we refer to the set of conditional quantum states $\{\rho_{a|x}\}_{a,x}$ ($\{\rho_{a|x}\}$ in short) as an \emph{assemblage}.

In certain cases, instead of the usual quantum mechanical description, the preparation of an assemblage $\{\rho_{a|x}\}$ can be understood via a semiclassical model. Specifically, following Ref.~\cite{wiseman2007}, we say that an assemblage $\{\rho_{a|x}\}$ admits a local-hidden-state (LHS) model if there exists legitimate probability distributions $P(\lambda)$, $P(a|x,\lambda)$, and normalized quantum states $\hat{\sigma}_\lambda$ such that
\begin{equation}
	\rho_{a|x} = \sum_\lambda P(a|x,\lambda)P(\lambda)\hat{\sigma}_\lambda \quad \forall\,\, a,x,
\label{Eq_LHS}
\end{equation}
i.e., the observed assemblage is an average of quantum states $\hat{\sigma}_\lambda$ distributed to Bob over the common-cause distribution $P(\lambda)$ and the local response function $P(a|x,\lambda)$ on Alice's end. In this case, it is conventional to refer to the assemblage as being {\em unsteerable}. Otherwise, an assemblage $\{\rho_{a|x}\}$ that cannot be decomposed in the form of Eq.~\eqref{Eq_LHS} is said to be {\em steerable}, as Alice can apparently {\em steer} the ensemble of quantum states at Bob's end with her choice of local measurements.

There are several ways to quantify the degree of steerability of any given assemblage $\{\rho_{a|x}\}$, e.g., the steerable weight~\cite{SNC14}, the steering robustness~\cite{Piani15}, the relative entropy of steering~\cite{Gallego15,Eneet17b}, the optimal steering fraction~\cite{HLL2016}, consistent trace-distance measure \cite{KuPRA18} etc. In this paper, we would focus predominantly on the steering robustness ($SR$), defined~\cite{Piani15} as the minimum (unnormalized) weight associated with another assemblage $\{\tau_{a|x}\}$ so that its mixture with $\{\rho_{a|x}\}$ is unsteerable, i.e.,\footnote{Throughout, we use $A\succeq B$ to mean that $A-B$ is positive semidefinite.}
\begin{equation}\label{Eq_DefineSR}
\begin{aligned}
{\rm SR}(\{\rho_{a|x}\}) := &\,\,\min_{t,\{\sigma_\lambda\}, \{\tau_{a|x}\}} \quad t\\
\text{s.t.}~
&\frac{\rho_{a|x} + t \tau_{a|x}}{1+t} = \sum_\lambda D(a|x,\lambda)\sigma_\lambda \quad \forall\,\, a,x,\\
& \sigma_\lambda\succeq0,\quad \sum_\lambda \tr(\sigma_\lambda)=1,\\
&\{\tau_{a|x}\}\ \ \text{is a valid assemblage},
\end{aligned}
\end{equation}
where $D(a|x,\lambda)=\delta_{a,\lambda_x}$, $\lambda=(\lambda_1,\ldots,\lambda_{n_x})$, and $\sigma_\lambda$ is a subnormalized quantum state [$\sigma_\lambda=P(\lambda)\hat{\sigma}_\lambda$, cf. Eq.~\eqref{Eq_LHS}]. In the above formulation, we have made use of the fact that, in determining the existence of a decomposition in the form of Eq.~\eqref{Eq_LHS}, it suffices to consider deterministic $P(a|x,\lambda)$ in the form just described.

A prominent advantage of SR is that, as with steerable weight~\cite{SNC14}, it can be efficiently computed as a semidefinite program (SDP) [by setting $(1+t)\sigma_\lambda$ as $\rho_\lambda$ in Eq.~\eqref{Eq_DefineSR}]:
\begin{subequations}
\begin{align}
{\rm SR}(\{\rho_{a|x}\}) = & \ \ \min_{\{\rho_\lambda\}} \ \ \sum_{\lambda}\tr\left(\rho_\lambda\right) - 1\label{Eq_SR1}\\
\text{s.t.}~
&\sum_\lambda D(a|x,\lambda)\rho_\lambda \succeq \rho_{a|x}  \quad \forall\  a,x,\label{Eq_SR2}\\
&\rho_\lambda \succeq 0 \quad \forall\ \lambda.\label{Eq_SR3}
\end{align}
\label{Eq_SR}
\end{subequations}
From the dual of this SDP (see Ref.~\cite{Cavalcanti17}), one finds that {\blu {SR}} actually coincides with the optimal steering fraction, a steering monotone (based on optimal steering inequalities) introduced in Ref.~\cite{HLL2016}. Finally, as remarked by Piani and Watrous~\cite{Piani15}, {\blu SR} can be given an operational meaning in terms of the (relative) success probability of some quantum information tasks (more on this below). 


\subsection{The framework of assemblage moment matrices}\label{Sec_AMMs_formulation}

In a DI setting, single-partite probability distributions $P(a|x)$, $P(b|y)$ {\em alone} cannot be used to provide nontrivial characterizations of the underlying devices. This is because for one to arrive at any nontrivial statement, the observed correlation $\Pobs$ must also violate a Bell inequality~\cite{Brunner08,Scarani12}. Since single-partite probability distributions alone do not reveal any correlation between the measurement outcomes of distant parties, they cannot possibly violate any Bell inequalities. Following this reasoning, it {\blu may seem  the} case that moment matrices associated with single-partite density matrices are also useless for DI characterizations.

While this intuition is  true for normalized single-partite density matrices, the same cannot be said when it comes to an assemblage, which consists only of subnormalized density matrices that arise in a steering experiment.
Specifically, for each combination of outcome $a$ and setting $x$, applying the local CP map of Eq.~\eqref{Eq:LocalMapping} to the conditional state $\rho_{a|x}$, cf. Eq.~\eqref{Eq_quantum_assemblage}, gives rise to a matrix of expectation values:
\begin{equation}
\begin{aligned}
	\chi[\rho_{a|x},\{B_i\}] &= \Lambda_{\text{B}}(\rho_{a|x})\\
	&= \sum_{ij}\ket{i}\!\bra{j}\tr[\rho_{a|x} B_j^\dagger B_i]\quad \forall\ a,x,
	\end{aligned}
\label{Eq_AMM}
\end{equation}
where $\{B_i\}$ are again operators formed from the product of $\{\id\}\cup\{E^\tB_{b|y}\}_{y,b}$.
When {\blu the set $\{B_i\}$ involves} operators that are at most $\ell$-fold product of Bob's POVM elements, the collection of matrices in Eq.~\eqref{Eq_AMM} are said~\cite{SLChen16} to be the \emph{assemblage moment matrices} (AMMs) of level $\ell$, and we denote each of them by $\chil[\rho_{a|x}]$.

Indeed, as with the moment matrices introduced in Sec.~\ref{Sec_MM}, all entries of $\Pobs$ can be identified with entries in these single-partite moment matrices. For example, by using Eq.~\eqref{Eq_quantum_assemblage} in Eq.~\eqref{Eq_AMM} and choosing an entry in $\chil[\rho_{a|x}]$ such that $B_i=B_j=B_j^2=E^\tB_{b|y}$ for some $b,y$ gives $\tr[\rho_{a|x} B_j^\dagger B_i]=P(a,b|x,y)$.
In a DI setting, neither the assemblage $\{\rho_{a|x}\}$ nor the measurement assemblage $\{E_{b|y}^\tB\}$ is known. Thus, apart from the few entries that can be estimated, each of these moment matrices is (largely) uncharacterized. Let us denote the corresponding AMM in this setting by $\chilbb[\rho_{a|x}]$ and the corresponding unknown entries collectively by $\{u_i^{(a,x)}\}$.\footnote{Although the known data in these AMMs are $\Pobs$, we shall write $\chiDI$ instead of $\chi^{(\ell)}[\mathbf{P}_\text{obs}]$ to emphasize that the underlying moment matrices are induced by an assemblage. }  The requirement that each $\chilbb[\rho_{a|x}]$  is a legitimate moment matrix, i.e., is in the form of Eq.~\eqref{Eq_AMM} while assuming Eq.~\eqref{Eq_quantum_assemblage},
then allows one to approximate algorithmically (from outside) the set of quantum correlations $\Q$, cf. Eq.~\eqref{Eq:Quantum}.
In addition, as with the moment matrices discussed in Sec.~\ref{Sec_MM}, in determining if some given $\Pobs$ is quantum realizable, we may assume that all $\{E_{b|y}^\tB\}$ correspond to those of projective measurements while the unobservable expectation values are real numbers (see Table~\ref{TB_AMM} for a summary of the various entries of $\chilbb[\rho_{a|x}]$).

\begin{table}[h!]
\centering

\caption{Elements of the moment matrix $\chiDI$ constructed from Eq.~\eqref{Eq_AMM} with the simplification that all measurements are described by orthogonal projectors.
}
\begin{tabular}{|c|c|}
\hline
elements & for  $B_j^\dagger B_i$ \\ \hline \hline
0 & containing $E_{b|y}^\tB E_{b'|y}^\tB$ with $b\neq b'$  \\ \hline
$P_\text{obs}(a,b|x,y)$
& being $E_{b|y}^\tB $ \\ \hline
unknown $u_i\in\mathbb{R}$
& otherwise\\ \hline
\end{tabular}\label{TB_AMM}
\end{table}

As an explicit example, consider the $\ell=1$ AMMs with $n_y=n_b=2$, i.e., where $\{B_i\}=\{\id,E_{1|1}^\text{B},E_{1|2}^\text{B}\}$. From Eq.~\eqref{Eq_AMM} we have that for each $a$ and $x$:
\begin{equation}
\begin{aligned}
&\chi^{(1)}[\rho_{a|x},\{B_i\}] =\\
&\begin{pmatrix}
\tr(\rho_{a|x}) & \tr(\rho_{a|x} E_{1|1}^\text{B}) & \tr(\rho_{a|x} E_{1|2}^\text{B})\\
\tr(\rho_{a|x} E_{1|1}^\text{B}) & \tr(\rho_{a|x} E_{1|1}^\text{B}) & \tr(\rho_{a|x}E_{1|1}^{\text{B}\dag} E_{1|2}^\text{B})\\
\tr(\rho_{a|x} E_{1|2}^\text{B}) & \tr(\rho_{a|x} E_{1|2}^{\text{B}\dag} E_{1|1}^\text{B})  & \tr(\rho_{a|x} E_{1|2}^\text{B})
\end{pmatrix}.
\end{aligned}
\label{Eq_AMM_level1}
\end{equation}
For DI characterizations, we then write this matrix (for a fixed value of $a$ and $x$) as:
\begin{equation}
\begin{aligned}
&\chiDIfirst =\\ 
&\begin{pmatrix}
P_\text{obs}(a|x) & P_\text{obs}(a,1|x,1) & P_\text{obs}(a,1|x,2)\\
P_\text{obs}(a,1|x,1) & P_\text{obs}(a,1|x,2) & u_1^{(a,x)}\\
P_\text{obs}(a,1|x,2) & u_1^{(a,x)}  & P_\text{obs}(a,1|x,2)
\end{pmatrix},
\end{aligned}
\label{Eq_DIAMM_level1}
\end{equation}
where we have made use of the simplification mentioned above and expressed the experimentally inaccessible expectation value as:
\begin{equation}
	\tr(\rho_{a|x} E_{1|2}^{\text{B}\dag} E_{1|1}^\text{B}) = \tr(\rho_{a|x}E_{1|1}^{\text{B}\dag} E_{1|2}^\text{B}) =u_1^{(a,x)},
\end{equation}
with $u_i^{(a,x)}\in\mathbb{R}$  (see Ref.~\cite{Moroder13}). 

\section{Device-independent applications}
\label{Sec:DI-App}

Having recalled from Ref.~\cite{SLChen16} the AMM framework, we are now in a position to further explore the framework for DI characterizations.

\subsection{Quantification of steerability}

As was already noted in our previous work~\cite{SLChen16}, a DI lower bound on SR forms the basis of a couple of DI applications based on the AMM framework. For completeness and for comparison with the improved lower bound that we shall present in Sec.~\ref{Sec_DIIR}, we now explain how a DI lower bound on SR can be obtained by relaxing the optimization problem given in Eq.~\eqref{Eq_SR}, as was proposed in Ref.~\cite{SLChen16}.

To this end, let us emphasize once again that in the DI paradigm, one does not assume any knowledge (e.g., the Hilbert space dimension) of quantum states $\rho_\lambda$ and $\rho_{a|x}$. However, if the constraints of Eq.~\eqref{Eq_SR} hold, it must be the case that even upon the application of the local CP map given in Eq.~\eqref{Eq_AMM}, the constraints---which demand the positivity of certain matrices---would still hold. At the same time, notice that each $\tr\left(\rho_\lambda\right)$ appearing in the objective function of Eq.~\eqref{Eq_SR} can still be identified as a specific entry, denoted by $\chi_{\mbox{\tiny DI}}^{(\ell)}[\rho_{\lambda}]_\text{tr}$ in the AMM. For example, in the AMM given in Eq.~\eqref{Eq_AMM_level1}, the trace of the underlying {\blu matrix $\rho_{a|x}$} is given by the upper-left entry of the matrix. Putting all these together, we thus see that a DI lower bound on SR can be obtained by solving the following SDP:
\begin{subequations}\label{Eq_relax_SR}
\begin{align}
\min_{\{u_v\}} ~~&\left(\sum_{\lambda}\chi_{\mbox{\tiny DI}}^{(\ell)}[\rho_{\lambda}]_\text{tr}\right)-1 \label{Eq_relax_SR1}\\
\text{s.t.}~~ &\sum_{\lambda}D(a|x,{\lambda})\chi_{\mbox{\tiny DI}}^{(\ell)}[\rho_{\lambda}]\succeq
\chiDI \quad\forall~a,x,\label{Eq_relax_SR2}\\
&\chi_{\mbox{\tiny DI}}^{(\ell)}[\rho_{\lambda}]\succeq 0\quad \forall ~\lambda,\label{Eq_relax_SR3}\\
&\sum_a\chiDI = \sum_a\chi_{\mbox{\tiny DI}}^{(\ell)}[\rho_{a|x'}] ~~\forall\, x\neq x',\label{Eq_relaxSR_nosig}\\
&\sum_a\chiDI_\text{tr}=1,\quad \chiDI\succeq 0 ~~\forall ~a,x\label{Eq_relaxSR_posi},\\
&P(a,b|x,y)=P_{\mbox{\tiny obs}}(a,b|x,y)\quad\forall\quad a,b,x,y.\label{Eq_prob_match}
\end{align}
\end{subequations}
As explained above, Eq.~\eqref{Eq_relax_SR2} and Eq.~\eqref{Eq_relax_SR3} follow  by applying the CP map of Eq.~\eqref{Eq_AMM} to the constraints of Eq.~\eqref{Eq_SR}. However, by themselves, physical constraints (including normalization, positivity and consistency) associated with the assemblage $\{\rho_{a|x}\}$ may be violated and thus have to be separately enforced in Eq.~(\ref{Eq_relaxSR_nosig}) and (\ref{Eq_relaxSR_posi}). Empirical observation enters at the level of observed correlation in Eq.~\eqref{Eq_prob_match}, i.e., by matching entries in the AMM with the empirical data summarized in $\Pobs$. Instead of Eq.~\eqref{Eq_prob_match}, a (weaker) lower bound can also be obtained by imposing an equality constraint of the form $\sum_{a,b,x,y} \beta^{x,y}_{a,b} P(a,b|x,y) = \hat{I}_{\vec{\beta}}$ where $\hat{I}_{\vec{\beta}}$ is the observed value of a certain Bell function specified by real coefficients $\beta^{x,y}_{a,b}$. {\blu Moreover, notice that if we have access to the observed probabilities $\Pobs$, the condition $\sum_a\chiDI_\text{tr}=1$ is automatically satisfied, as it amounts to the condition $\sum_a P(a|x)=1$. On the other hand, this is not the case if we have access only to the Bell function  $\hat{I}_{\vec{\beta}}$.}

Importantly, the constraints of Eq.~\eqref{Eq_relaxSR_nosig} and Eq.~\eqref{Eq_relaxSR_posi} do not necessarily single out $\{\rho_{a|x}\}$ as the underlying assemblage; neither do Eq.~\eqref{Eq_relax_SR2} and Eq.~\eqref{Eq_relax_SR3} entail the constraints of Eq.~\eqref{Eq_SR}. The above optimization problem is thus a relaxation of that given in Eq.~\eqref{Eq_SR}. For concreteness, let us denote the optimum of Eq.~\eqref{Eq_relax_SR} by $\SRDIPobs$ and that obtained for some observed Bell violation as SR$_{\text{\tiny DI},\ell}^{\mbox{\tiny A$\rightarrow$B}}(\hat{I})$,
then 
\begin{equation}\label{Eq:SRBounds}
	{\rm SR}(\{\rho_{a|x}\}) \ge \SRDIPobs  \ge \text{SR}_{\text{\tiny DI},\ell}^{\mbox{\tiny A$\rightarrow$B}}(\hat{I})
\end{equation}
for all $\ell\ge 1$, thus giving the desired DI lower bound on ${\rm SR}(\{\rho_{a|x}\})$ [see Sec.~\ref{Sec:CS16} and also Ref.~\cite{Cavalcanti16} for alternative approaches for bounding ${\rm SR}(\{\rho_{a|x}\})$.]


\subsection{Quantification of the advantage of quantum states in subchannel discriminations}


From Eq.~\eqref{Eq:SRBounds}, one can also quantitatively estimate the usefulness of certain steerable quantum states in a kind of subchannel discrimination problem (see Ref.~\cite{Piani15} and references therein). To this end, let $\hat{\Lambda}=\sum_a \Lambda_a$ be a quantum channel (a trace-preserving CP map) that can be decomposed into a collection of subchannels $\{\Lambda_a\}_a$, i.e., a family of CP maps $\Lambda_a$ that are each trace nonincreasing for all input states $\rho$. Following Ref.~\cite{Piani15}, we refer to this collection of subchannels as a {\em quantum instrument} $\I=\{\Lambda_a\}_a$.  An example of $\I$ consists in performing measurement on the input state with ``$a$'' labeling the measurement outcome. 

In its primitive form, a subchannel discrimination problem concerns the following task: input a quantum state $\rho$ into the channel $\hat{\Lambda}$ and determine, for each trial, the actual evolution (described by $\Lambda_a$) that $\rho$ undergoes by performing a measurement on $\Lambda_a[\rho]$. For an input quantum state $\rho$, if we denote by $\{G_{a}\}_a$ the POVM associated with the measurement on the output of the channel, then the probability of correctly identifying the subchannel $\Lambda_a$ is given by 
\begin{equation}\label{Eq:ProbCorrect}
	p_\checkmark(\I, \{G_{a'}\}_{a'}, \rho) := \sum_a \tr\left(G_a\Lambda_a[\rho]\right).
\end{equation}
For any given quantum instrument $\I$, the maximal probability of correctly identifying the subchannel is then obtained by maximizing the above expression over the input state $\rho$ and the POVM $\{G_{a'}\}_{a'}$, i.e., 
\begin{equation}\label{Eq:ProbNE}
	p_\checkmark^{\text{NE}}(\I):=\max_{\rho,\{G_{a'}\}_{a'}}p_\checkmark(\I, \{G_{a'}\}_{a'}, \rho),
\end{equation}
where we use NE to signify ``no entanglement'' in the above guessing probability expression.

In Refs.~\cite{Piani09,Piani15}, the authors considered a situation where the input to the channel is a part of an entangled state $\rab$ (B is the part that enters the channel) and where a measurement on the output $\mathbb{I}_\tA\otimes\Lambda^\tB_a[\rab]$ is allowed. Suppose now that the final measurement is restricted to be separable across A and B, but allowed to be coordinated by one-way classical communication~\cite{Piani15} (one-way LOCC) from B to A, i.e., taking the form of
\begin{equation}\label{Eq:1-wayLOCC}
	G_{a'} = \sum_x E_{a'|x}^{\tA}\otimes E_x^{\tB},\\
\end{equation}
where $E_{a'|x}^{\tA}\succeq 0$, $\sum_{a'}E_{a'|x}^{\tA}=\id_\tA$ and $E_x^{\tB}\succeq 0$, $\sum_x E_x^{\tB}=\id_\tB$. Then, it was shown~\cite{Piani15} that for any steerable quantum state $\rab$, there always exists an instrument $\I=\{\Lambda_a\}_a$ such that the corresponding guessing probability---after optimizing over measurements of the form given in Eq.~\eqref{Eq:1-wayLOCC}---exceeds $p_\checkmark^{\text{NE}}(\I)$.

More precisely, let $\{G_{a'}\}_{a'}$ take the form of Eq.~\eqref{Eq:1-wayLOCC}. Then, for the initial state $\rab$, the corresponding guessing probability  (after optimizing over such measurements) is
\begin{equation}\label{Eq:Prob1way}
	p^{\mbox{\tiny B$\rightarrow$A}}_\checkmark(\I,\rab ):= \max_{\{G_{a'}\}_{a'}} \sum_a \tr\left(G_a\mathbb{I}_\tA\otimes\Lambda^\tB_a[\rab]\right).
\end{equation}
The advantage of a steerable state $\rab$ compared to unentangled resources in the subchannel discrimination task can then be quantified via the ratio of their success probabilities. In Ref.~\cite{Piani15}, this ratio was shown to be closely related to the ${\rm SR}^{\mbox{\tiny A$\rightarrow$B}}(\rab)$, the steering robustness of the given \emph{quantum state} $\rab$, defined as:
\begin{equation}
	{\rm SR}^{\mbox{\tiny A$\rightarrow$B}}(\rab):=\sup_{\{E_{a|x}^\text{A}\}}\SRrax.
\label{Eq_SRofState}
\end{equation}
Explicitly, since~\cite{Piani15} 
\begin{equation}\label{Eq:SRtoQITask}
	\sup_\I\frac{p^{\mbox{\tiny B$\rightarrow$A}}_\checkmark(\I, \rab)}{p_\checkmark^{\text{NE}}(\I)} = {\rm SR}^{\mbox{\tiny A$\rightarrow$B}}(\rab)+1,
\end{equation}
and we can provide a DI lower bound on $\SRrax$ via Eq.~\eqref{Eq:SRBounds}, it follows from Eq.~\eqref{Eq_SRofState} that we can also estimate in a DI manner the advantage of the measured state over unentangled resources for the task of subchannel discrimination.

\subsection{Quantification of entanglement}
\label{Sec:DI-ER}

The possibility to lower bound the entanglement of an underlying state in a DI setting was first demonstrated---using negativity~\cite{Vidal02} as the entanglement measure---in Ref.~\cite{Moroder13}. Subsequently, in Ref.~\cite{Toth15}, this possibility was extended to include the linear entropy of entanglement. In this subsection, we discuss how such a quantification can be achieved also for the generalized robustness of entanglement~\cite{vidal1999,Steiner03} defined as:
\begin{equation}
\begin{aligned}
	\ER(\rab):= \min_{t, \tab} & ~~t\geq 0\\
	\text{s.t.}& ~~\frac{\rab + t \tab}{1+t}\quad \text{is separable},\\
	& ~~\tab \quad \text{is a quantum state}.
\end{aligned}
\label{Eq_ER}
\end{equation}


\subsubsection{Via the approach of AMM}

To obtain a DI lower bound on $ER$, we first remind that the set of unsteerable states (either from A to B, or from B to A) is a strict superset to the set of separable states. Hence, it is evident from Eq.~\eqref{Eq_ER} that (see also Ref.~\cite{Piani15})
\begin{equation}
	\ER(\rab)\geq {\rm SR}(\rab):=\max\{{\rm SR}^{\mbox{\tiny A$\rightarrow$B}}(\rab),{\rm SR}^{\mbox{\tiny B$\rightarrow$A}}(\rab)\}.
\label{Eq_ERgeqSR}
\end{equation}
It then immediately follows from Eq.~\eqref{Eq:SRBounds} and Eq.~\eqref{Eq_SRofState} that for any assemblage on {\blu Bob's} side $\{\rho_{a|x}\}$, any assemblage on {\blu Alice's} side $\{\rho_{b|y}\}$, or any correlation $\Pobs$ associated with these assemblages observed in a Bell experiment:
\begin{equation}\label{Eq_ERgeqSRDI}
\begin{split}
	\ER(\rab)&\geq \max\{{\rm SR}(\{\rho_{a|x}\}),{\rm SR}(\{\rho_{b|y}\})\},\\
		    &\ge \max\{\SRDIPobs,{\rm SR}_{\text{\tiny DI},\ell}^{\mbox{\tiny B$\rightarrow$A}}(\Pobs)\},
\end{split}
\end{equation}
which give the desired DI lower bounds on ER$(\rab)$.

\subsubsection{Via the approach of nonlocal robustness}\label{Sec:CS16}

In Ref.~\cite{Cavalcanti16}, Cavalcanti and Skrzypczyk introduced, for any given correlation $\{P(a,b|x,y)\}$, a quantifier for nonlocality by the name of \emph{nonlocal robustness}:
\begin{equation}\label{Eq_NR}
\begin{aligned}
	\NR(\vecP):= \min_{r, \{ Q(a,b|x,y)\}} &\quad r\geq 0\\
	\text{s.t.}& \quad\Big\{ \frac{P(a,b|x,y) + r Q(a,b|x,y)}{1+r} \Big\} \in \mathcal{L}\\
	& \quad\{Q(a,b|x,y)\}\in \mathcal{Q},
\end{aligned}
\end{equation}
where $\mathcal{L}$ and $\mathcal{Q}$ are, respectively, {\blu the sets } of Bell-local and quantum correlations. Moreover, they~\cite{Cavalcanti16} showed that the nonlocal robustness $\NRPabxy$ for any correlation associated with an assemblage is a lower bound on the corresponding steering robustness, i.e., 
\begin{equation}
{\rm SR}(\{\rho_{a|x}\})\geq\NRPabxy.
\label{Eq_SRgeqNR}
\end{equation}
Hence, by using the first inequality of Eq.~\eqref{Eq_ERgeqSR}, we see that a DI lower bound on $\ER(\rab)$ can also be obtained by computing $\NR(\Pobs)$.

\subsubsection{Via an MBLHG-based~\cite{Moroder13} approach}

For comparison, let us mention here also the possibility for bounding $\ER(\rab)$ based on the approach of Moroder \emph{et al.}~\cite{Moroder13}, abbreviated as MBLHG (see Sec.~\ref{Sec_MM}).
The idea is to first relax the separability constraint of Eq.~\eqref{Eq_ER} by the positive-partial-transposition constraint~\cite{Peres96,Horodecki96}, thereby making the optimum of 
the following SDP, i.e., 
\begin{equation}
\begin{aligned}
\min_{\omega_{\mbox{\tiny AB}}}& \quad\tr(\oab)-1\\
\text{s.t.}& \quad \omega_{\mbox{\tiny AB}}^{\mbox{\tiny T}_A}\succeq 0,
 \quad \oab \succeq \rab,
\end{aligned}
\label{Eq_ER_SDP}
\end{equation}
a lower bound on $\ER(\rab)$; here, we use $O^{\mbox{\tiny T}_A}$ to denote the partial transposition of operator $O$ with respect to the Hilbert space of A. For a two-qubit state or a qubit-qutrit state $\rab$, the result of Horodecki~{\em et al.}~\cite{Horodecki96}  implies that the $\ER(\rab)$ computed from Eq.~\eqref{Eq_ER_SDP} is tight.

Next, by applying the local mapping of Eq.~\eqref{Eq:LocalMapping} to the linear matrix inequality constraints of Eq.~\eqref{Eq_ER_SDP}, we obtain a further relaxation of  Eq.~\eqref{Eq_ER}---and hence also a DI lower bound on $\ER(\rab)$---by solving the following SDP:
\begin{equation}
\begin{aligned}
\min_{\chi[\oab],\{u_i\}}& \quad\chi[\oab]_{\tr} -1\\
\text{s.t.}& \quad \chi[\oab]^{\mbox{\tiny T}_{\bar{A}}}\succeq 0,
 \quad \chi[\oab] \succeq \chi[\rab],\\
& \quad \chi[\oab] \succeq 0,\quad \chi[\rab] \succeq 0,\quad \chi[\rab]_{\tr} = 1,\\
&P(a,b|x,y)=P_{\mbox{\tiny obs}}(a,b|x,y)\quad\forall\quad a,b,x,y,
\end{aligned}
\label{Eq_DIER_MM}
\end{equation}
where $\chi[.]$ refers to moment matrix in the form of Eq.~\eqref{Eq_MM1}, $\{u_i\}$ is the set of unknown moments in $\chi[\rab]$, and the empirical observation enters, as with Eq.~\eqref{Eq_prob_match}, by imposing the last line of equality constraints for the relevant entries in $\chi[\rab]$. Note also that the second line of constraints on $\chi[\rab]$ stems from the fact that we now no  longer assume anything about the underlying state $\rab$, but only constraints of the form of Eq.~\eqref{Eq_prob_match}. Hereafter, we  denote the optimum of Eq.~\eqref{Eq_DIER_MM} by $\ER_{\text{\tiny DI},\ell}(\Pobs)$.

In Table~\ref{TB_DIER}, we summarize how a DI lower bound on $\ER(\rab)$ can be obtained using the three approaches explained above.
\begin{center}
\begin{table}[h!]
\centering
\caption{Different approaches to DI quantification of the {\blu generalized} robustness of entanglement. Here and below, we use $^\dag$ to point out a new method introduced in the present work for bounding quantities of interest in a DI manner.
}
\begin{tabular}{|c|c|}
\hline
method & bound relations  \\ \hline \hline
MBLHG-based~\cite{Moroder13}$^\dag$
& $\ER(\rab)\geq \ER_{\text{\tiny DI},\ell}[\mathbf{P}_{\text{obs}}]$ \\ \hline
CS~\cite{Cavalcanti16}
& $\ER(\rab)\geq {\rm SR}(\{\rho_{a|x}\})\geq \NRPobs$ \\ \hline
CBLC~\cite{SLChen16}
& $\ER(\rab)\geq {\rm SR}(\{\rho_{a|x}\})\geq \SRDIPobs $\\ \hline

\end{tabular}\label{TB_DIER}
\end{table}
\end{center}

\subsubsection{Some explicit examples}

To gain some insight on the tightness of the DI bounds provided by the aforementioned approaches, consider, for example, the isotropic states~\cite{Horodecki99}:
\begin{equation}\label{Eq:IsoStates}
	\rIdv = v_d\proj{\Phi^+_d} + (1-v_d)\frac{\id}{d^2},\quad  -\frac{1}{d^2-1}\leq v_d \leq 1,
\end{equation}
where $\ket{\Phi^+_d}=\frac{1}{\sqrt{d}}\sum_{i=1}^d \ket{i}\ket{i}$ is the $d$-dimensional maximally entangled state, and $\frac{\id}{d^2}$ is the two-qudit maximally mixed state. It is known that these states are entangled if and only if $v_d>\frac{1}{d+1}$.

\begin{figure}[h!]
\begin{center}
\emph{\includegraphics[width=9cm]{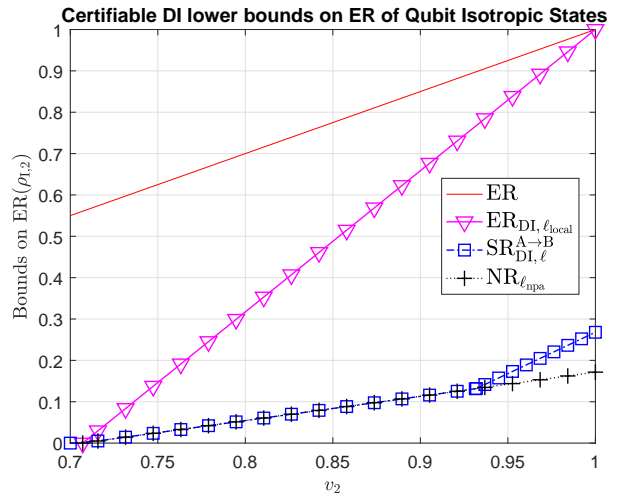} }
\caption{\label{Fig:ERBoundsSimplified}
Certifiable DI lower bounds on the generalized robustness of entanglement ($\ER$) for two-qubit isotropic states $\rho_{\text{\tiny I}, 2}(v_2)$ based on various Bell-inequality-violating correlations $\vecP$ obtained from these states using the three approaches discussed in Sec.~\ref{Sec:DI-ER} (see text and Fig.~\ref{Fig:FullDetails} for further details). Bounds obtained from the approach of MBLHG~\cite{Moroder13}, AMM~\cite{SLChen16}, and of Ref.~\cite{Cavalcanti16} are marked, respectively, using  triangles ({\color{magenta} $\triangledown$}), squares ({\color{blue} $\square$}) , and crosses ($+$). For completeness, the actual value of $\ER[\rho_{\text{\tiny I}, 2}(v_2)]$ for each given value of visibility $v_2$, cf.  Eq.~\eqref{Eq:ER-rhoIso}, is also included as a (red) solid line.  
}
\label{Fig_comparison_NR_and_DISR_Werner_state}
\end{center}
\end{figure}

In Appendix~\ref{App:ERStates}, we show that the generalized robustness of entanglement for these states are:
\begin{equation}\label{Eq:ER-rhoIso}
	\ER[\rIdv]=\max\left\{0,\frac{d-1}{d}\left[(d+1)v_d-1\right]\right\}.
\end{equation}
To compare the efficiency of these three methods in lower bounding $\ER[\rIdv]$ in a DI setting, we first consider $\rho_{\text{\tiny I},2}$ in conjunction with their optimal measurements with respect to the Clauser-Horne-Shimony-Holt (CHSH) Bell inequality~\cite{Clauser69} (see, e.g., Chapter 6 and Appendix B.4.1 of Ref.~\cite{Liang:PhDthesis}), the $I_{3322}$~\cite{Collins04_0} Bell inequality, and the elegant Bell inequality~\cite{Gisin:ManyQuestions} (see, e.g., Ref.~\cite{Christensen15}), respectively. The correlation $\vecP=\{P(a,b|x,y)\}$ obtained therefrom for each of these Bell scenarios is then fed into the SDP of Eq.~\eqref{Eq_relax_SR}, Eq.~\eqref{Eq_NR} and Eq.~\eqref{Eq_DIER_MM}, respectively, to obtain the corresponding DI lower bound on $\ER[\rIdv]$ (cf. Table~\ref{TB_DIER}). The {\em best} lower bounds obtainable for each approach are shown in Fig.~\ref{Fig_comparison_NR_and_DISR_Werner_state} (for the lower bounds obtained for each approach in each Bell scenario, see Fig.~\ref{Fig:FullDetails}).


For visibilities less than $v_2\approx 0.9314$, the lower bounds obtained from the approach of AMM~\cite{SLChen16} and that of Ref.~\cite{Cavalcanti16} seem to fit well with the expression $(\sqrt{2}v_2-1)(\sqrt{2}-1)$. 
But for greater values of $v_2$, especially for $v_2\gtrsim0.9321$, the AMM-based lower bounds appear to be somewhat tighter, and appear to fit nicely with the expression $2v_2-\sqrt{3}$. On the other hand, it is also clear from the Figure that the lower bounds $\ER_{\text{\tiny DI},\ell}$ offered by the MBLHG-bsaed approach---which are well-represented by the expression\footnote{In general, since the correlations $\vecP$ employed for a particular value of $v_2\in[\tfrac{1}{\sqrt{2}},1]$ is a convex combination of the $\vecP$ for $v_2\approx\tfrac{1}{\sqrt{2}}$ and $v_2=1$, the DI bounds on $\ER[\rIdv]$ can be shown to be a convex function of $v_2$.}  $\frac{\sqrt{2}v_2-1}{\sqrt{2}-1}$---considerably outperform the lower bounds obtained from the other two approaches. 

\begin{figure}[h!]
\begin{center}
\emph{\includegraphics[width=9cm]{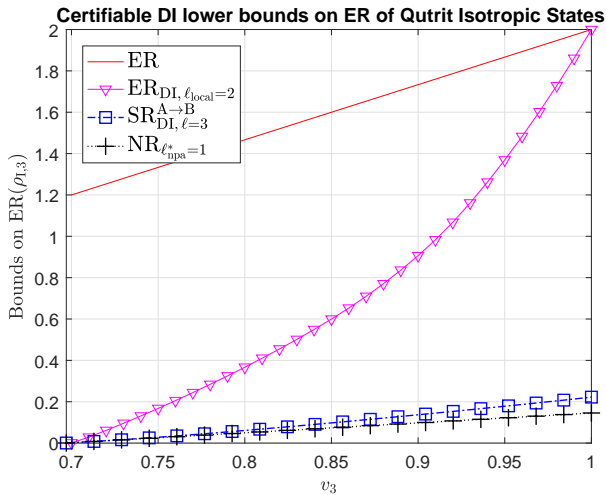} }
\caption{
Certifiable DI lower bounds on the generalized robustness of entanglement ($\ER$) for two-qutrit isotropic states $\rho_{\text{\tiny I}, 3}(v_3)$ based on the maximum $I_{2233}$-Bell-inequality-violating correlations $\vecP$ obtained from these states using the three approaches discussed in Sec.~\ref{Sec:DI-ER}. Legends of the figure follow that of Fig.~\ref{Fig:ERBoundsSimplified}. $\ell$ in the legend denotes the level of the SDP hierarchy involved in the computation; a $*$ is included as a superscript of $\ell$ whenever the next level of the hierarchy, $\ell+1$, gives the same SDP bound (within a numerical precision of the order of $10^{-6}$.
}
\label{Fig_comparison_NR_DISR_qutrit_states_2233scenario}
\end{center}
\end{figure}

As a second example, we consider the $d=3$ case of Eq.~\eqref{Eq:IsoStates} and the correlations leading to the optimal quantum violation of the $I_{2233}$-Bell inequality~\cite{Collins04_0} by these states. Our results are shown in Fig.~\ref{Fig_comparison_NR_DISR_qutrit_states_2233scenario}. Again, as with the case shown in Fig,~\ref{Fig_comparison_NR_DISR_qutrit_states_2233scenario}, the AMM approach appears to offer a somewhat tighter lower bounds than that of Ref.~\cite{Cavalcanti16}. Also, the MBLHG-based approach again appears to give a much better lower bound on $\ER[\rho_{\text{\tiny I}}(V_3)]$ than the other two approaches.

\subsection{Quantification of measurement incompatibility}\label{Sec_DIIR}



A collection of measurement, i.e., a measurement assemblage~\cite{Piani15} $\{E_{a|x}\}_{a,x}$ with $a$ denoting the output and $x$ the input, is said to be incompatible (not jointly-measurable) whenever it cannot be written in the form
\begin{equation}
	E_{a|x} = \sum_\lambda D(a,x|\lambda) G_\lambda,\ \forall \ a,x,
\end{equation}
where $G_\lambda \succeq 0$, $\sum_\lambda G_\lambda =\id$, and $D(a|x,\lambda)$  can be chosen, without loss of generality, as $D(a|x,\lambda) =\delta_{a,\lambda_x}$ [cf. Eq.~\eqref{Eq_DefineSR} and the text thereafter]. In other words, a measurement assemblage is  incompatible if there does not exist a joint measurement $\{G_\lambda\}$ providing all the outcome probabilities for any input. 

The use of incompatible measurement is necessary to observe both nonlocality \cite{Wolf09} and steering \cite{Uola14,Quint14}. Moreover, steering and incompatibilty problems can be mapped from one into another \cite{Uola15}, thus suggesting a measure of incompatibility, the incompatibility robustness (IR) introduced in Ref.~\cite{Uola15}. In analogy to the steering robustness, IR may be computed by solving the following SDP:
\begin{equation}\label{eq:IR_SDP}
\begin{aligned}
&{\rm IR}(\{E_{a|x}\}) = \min_{\{\tGl\}} \frac{1}{d}\sum_\lambda \tr[\tGl]-1\\
\text{s.t.}~
&\sum_\lambda D(a|x,\lambda)\tGl \succeq E_{a|x} \quad \forall\,\, a,x,\\
& \tGl \succeq 0 \quad \forall\,\, \lambda,\\
&\sum_\lambda \tGl = \openone \frac{1}{d}\sum_\lambda \tr[\tGl].
\end{aligned}
\end{equation}
where $d$ is the dimension of $\{E_{a|x}\}$. In Ref.~\cite{SLChen16}, it has been proven that the steering robustness of a given assemblage $\{\rho_{a|x}\}$ is a lower bound on the incompatibility robustness of the steering equivalent observables \cite{Uola15} $B_{a|x}= \rho_B^{-\frac{1}{2}} \rho_{a|x} \rho_B^{-\frac{1}{2}}$
with $\rho_B=\sum_a \rho_{a|x}$,\footnote{In the case of a reduced state $\rho_B$ not of full rank, it is sufficient to project the observables to its range, as discussed in Ref.~\cite{Uola15}. The same reasoning applies to the mapping of the two SDPs below.} which, in turn, is a lower bound on the incompatibility robustness of $\{E_{a|x}\}$, namely
\begin{equation}
{\rm IR}(\{E_{a|x}\})\geq {\rm IR}(\{B_{a|x}\})\geq {\rm SR}(\{\rho_{a|x}\}).
\end{equation}
The corresponding DI quantifier has then been discussed in Ref.~\cite{SLChen16}. An analogous observation has been made in Ref.~\cite{Cavalcanti16}, where Cavalcanti and Skrzypczyk also gave a lower bound on the degree of incompatibility, quantified by the incompatibility robustness of Alice's measurement assemblage $\{E_{a|x}^A\}$ in a DI manner. In their work, they first introduced a modified quantifier of steerability, called the \emph{consistent steering robustness}, defined as:
\begin{equation}\label{Eq:SRc}
\begin{aligned}
&{\rm SR}^c(\{\rho_{a|x}\}) = \min_{t,\{\tau_{a|x}\},\{\sigma_\lambda\}} \quad t \geq 0\\
\text{s.t.}~
&\frac{\rho_{a|x} + t \tau_{a|x}}{1+t} = \sum_\lambda D(a|x,\lambda)\sigma_\lambda \quad \forall\,\, a,x,\\
&\{\tau_{a|x}\}\quad\text{is a valid assemblage},\\
&\sigma_\lambda \succeq 0 \quad \forall\,\, \lambda,\quad \sum_\lambda \tr(\sigma_\lambda) = 1,\\
&\sum_a\tau_{a|x} = \sum_a\rho_{a|x}\quad\forall\,\, x.
\end{aligned}
\end{equation}
Compared with Eq.~\eqref{Eq_DefineSR}, the consistent steering robustness needs more constraints, i.e., $\sum_a\tau_{a|x}=\sum_a\rho_{a|x}$ for all $x$. The above problem can also be formulated as the following SDP [by setting $\ts_\lambda=(1+t)\sigma_\lambda$ and noting the non-negativity of $\tau_{a|x}$]:
\begin{equation}\label{Eq_SR_c}
\begin{aligned}
&{\rm SR}^c(\{\rho_{a|x}\}) = \min_{\{\ts_\lambda\}}\quad \tr\sum_{\lambda}\ts_\lambda - 1\\
\text{s.t.}~
&\sum_\lambda D(a|x,\lambda)\ts_\lambda \succeq \rho_{a|x}  \quad \forall\,\, a,x,\\
&\ts_\lambda \succeq 0 \quad \forall\,\, \lambda,\\
&\sum_\lambda \ts_\lambda = \tr\Big[\sum_\lambda \ts_\lambda\Big] \cdot \sum_a\rho_{a|x} ~~\forall\,\, x.
\end{aligned}
\end{equation}
Following an argument analogous to those in Ref.~\cite{SLChen16}, one can straightforwardly prove that ${{\rm SR}^c(\{\rho_{a|x}\}) = {\rm IR}(\{B_{a|x}\}})$ for the steering equivalent observables $\{B_{a|x}\}$.  In fact, by a direct inspection of Eqs.~\eqref{eq:IR_SDP} and \eqref{Eq_SR_c}, one sees that the SDP for computing ${\rm IR}(\{B_{a|x}\})$, cf.~Eq.~\eqref{eq:IR_SDP}, can be transformed into the one for computing ${\rm SR}^c(\{\rho_{a|x}\}$, Eq.~\eqref{Eq_SR_c}, via the mappings $E_{a|x} \mapsto {B_{a|x} =  \rho_B^{-\frac{1}{2}} \rho_{a|x} \rho_B^{-\frac{1}{2}}}$,  ${\tGl = \rho_B^{-\frac{1}{2}} \ts_\lambda \rho_B^{-\frac{1}{2}}}$, and the fact that $\sum_{a} \rho_{a|x} =\rho_B$. To show the inverse transformation,it is sufficient to use the inverse of the above mappings.

In order to provide a DI lower bound on ${\rm SR}^c(\{\rho_{a|x}\}$, the authors of Ref.~\cite{Cavalcanti16} introduced a nonlocality quantifier [for a given correlation $\vecP$] {\blu named} \emph{consistent nonlocal robustness} $\NRcPabxy$:
\begin{equation}
\begin{aligned}
&\NRcPabxy = \min_{r,\{Q(a,b|x,y)\}} \quad r \geq 0\\
\text{s.t.}~
&\frac{P(a,b|x,y) + r Q(a,b|x,y)}{1+r} \\&= \sum_\lambda D(a|x,\lambda)D(b|x,\lambda)P(\lambda) \quad \forall\,\, a,b,x,y,\\
&\{Q(a,b|x,y)\}\in\mathcal{Q},\\
&Q(b|y)=P(b|y)\quad\forall\,\, b,y,
\label{Eq_DefineNRc}
\end{aligned}
\end{equation}
i.e., it calculates the minimum noise one has to mix into $\{P(a,b|x,y)\}$ to make the mixture become local. $\{Q(a,b|x,y)\}\in\mathcal{Q}$ denotes $\{Q(a,b|x,y)\}$ that has a quantum realization, cf. Eq.~\eqref{Eq:Quantum}, and the last set of constrains requires the equivalence between the marginals of $\{P(a,b|x,y)\}$ and $\{Q(a,b|x,y)\}$, similar to the case of the consistent steering robustness [see the last line of Eq.~\eqref{Eq:SRc}]. Since the quantum set $\mathcal{Q}$ is not easily characterized, one can rather consider a superset $\tilde{\mathcal{Q}}^{(\ell)}$ of $\mathcal{Q}$ by using the $\ell$-th level of NPA hierarchy. In this way, one obtains a lower bound on $\NRc(\{P(a,b|x,y)\})$ by solving the following SDP, which is reformulated from Eq.~\eqref{Eq_DefineNRc} [by setting $q(\lambda)=\tfrac{1+r}{r}P(\lambda)$]:
\begin{equation}
\begin{aligned}
&\NRc_\ell(\{P(a,b|x,y)\}) = 1/s^*, \text{ with }\\
s^* = & \max_{\{q(\lambda)\},s} s\\
\text{s.t.}~~&s={\sum_\lambda q(\lambda)-1},\ s\geq 0,\\
\Bigg\{ &\sum_\lambda D(a|x,\lambda)D(b|y,\lambda)q(\lambda) - \\
&\left(\sum_\lambda q(\lambda)-1\right)\cdot P(a,b|x,y) \Bigg\}\in\tilde{\mathcal{Q}}^{(\ell)},\\
&\sum_\lambda D(b|y,\lambda)q(\lambda)=P(b|y)\cdot\sum_\lambda q(\lambda)\quad\forall\,\, b,y,\\
&q(\lambda) \geq 0 \quad\forall\,\, \lambda.
\end{aligned}
\label{LP_NRc}
\end{equation}

Using the above quantifiers, Cavalcanti and Skrzypczyk proved \cite{Cavalcanti16}
\begin{equation}
{\rm IR}(\{E_{a|x}\})\geq {\rm SR}^c(\{\rho_{a|x}\})\geq \NRc(\{P(a,b|x,y)\}),
\label{IR_SRc_NRc}
\end{equation}
which allows one to estimate the degree of incompatibility of Alice's measurements from the observed data $\Pobs$, i.e., in a DI manner. 

\begin{figure*}
\begin{minipage}[c]{.49\textwidth}
\emph{\includegraphics[width=8cm]{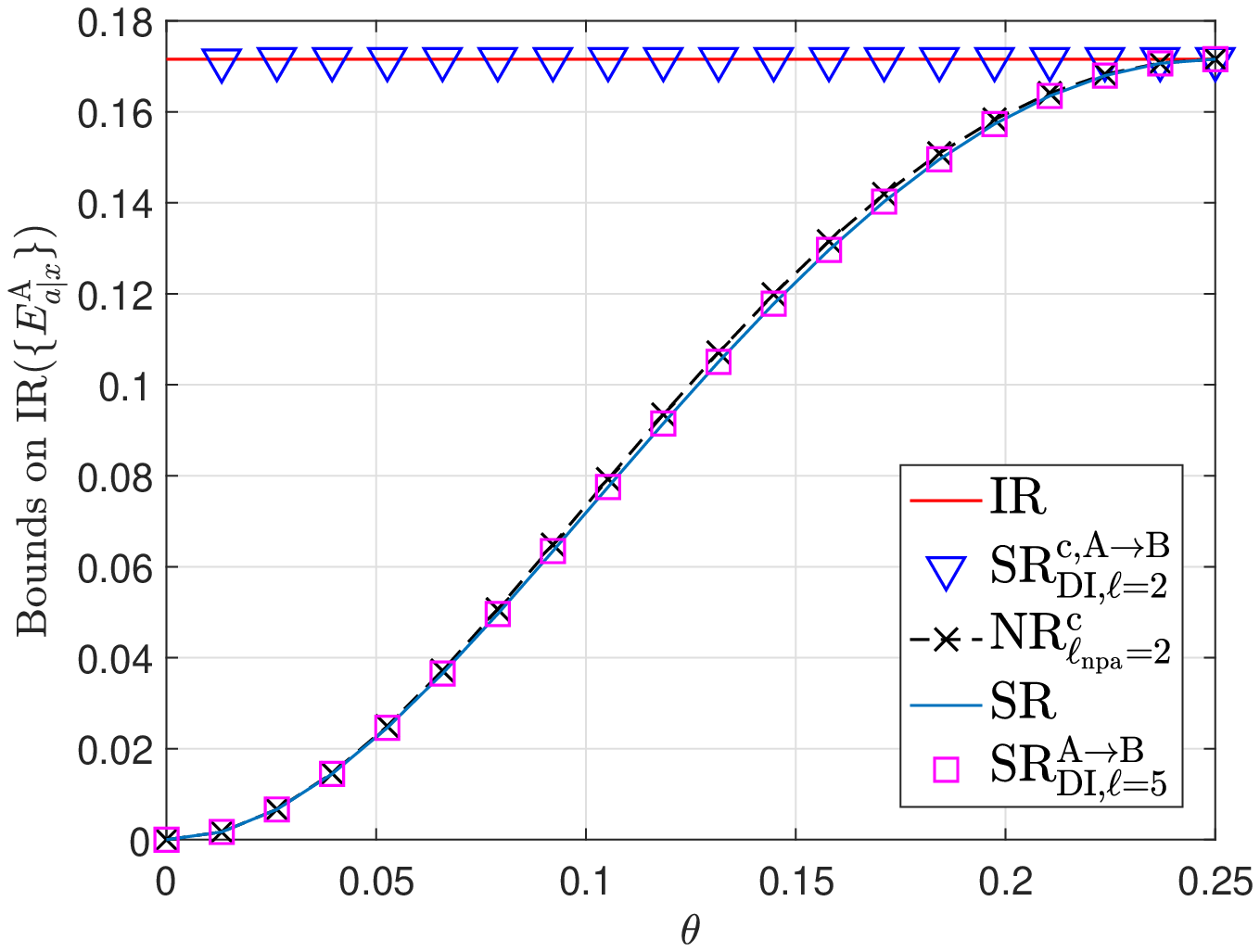} }
\text{(a) For Alice's measurement assemblage}
\label{Fig_comparison_DIIR0_Alice}
\end{minipage}
\begin{minipage}[c]{.49\textwidth}
\emph{\includegraphics[width=8cm]{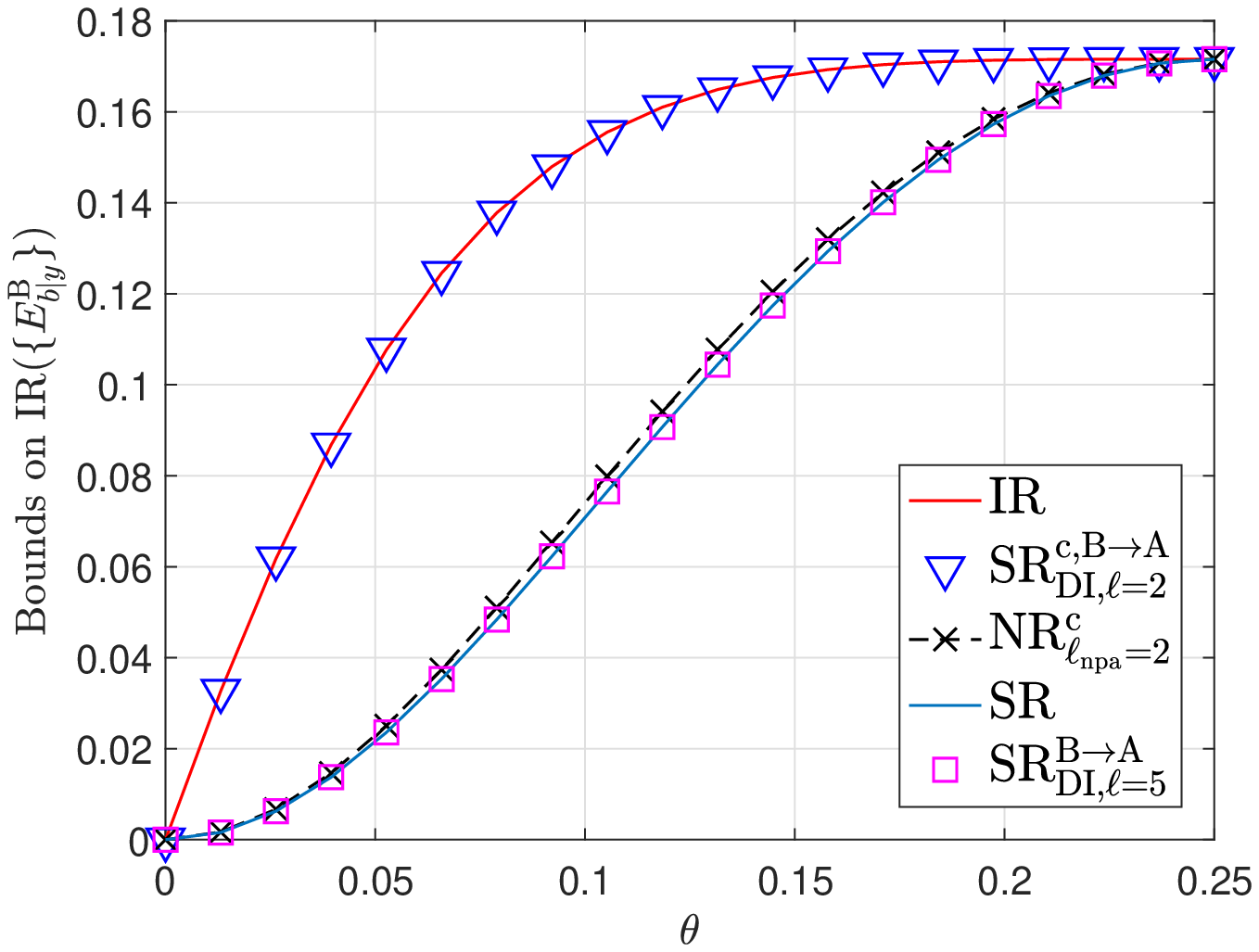} }
\text{(b) For Bob's measurement assemblage}
\label{Fig_comparison_DIIR_BtoA.eps}
\end{minipage}
\caption{\label{Fig:IR}
Comparison of DI lower bounds on measurement incompatibility---as measured by the incompatibility robustness $\IR$---of the measurements employed in attaining the optimal Bell-CH inequality violation of pure (partially) entangled  two-qubit states. The $\IR$ of the optimal measurement assemblage as a function of $\theta$ [cf. Eq.~\eqref{Eq_pure_entangled_state}] is marked with a (red) solid line. Following Ref.~\cite{Liang:PhDthesis}, we take the optimal measurements on Alice's side to be $\sigma_x$ and $\sigma_z$ [independent of $\theta$, see subplot (a)] while those on Bob's side to be a pair of measurements that are orthogonal on the Bloch sphere at $\theta=\frac{\pi}{4}$, but which gradually become aligned as $\theta$ decreases to 0 [see subplot (b)]. From the resulting optimal correlations $\vecP$, one can estimate, in a DI manner, $\IREax$ or $\IREby$ given by the AMMs approach ({\color{magenta} $\square$}), the Cavalcanti-Skrzypczyk approach ($\times$)~\cite{Cavalcanti16}, and the improved AMMs approach ({\color{blue} $\triangledown$}) introduced in this work. For comparison, we have also included the actual value of $\IR$ and SR in each plot using, respectively, a red (upper) and a turqoise (bottom) solid line.
}
\end{figure*}

Here, we would like to compare our method of lower-bounding incompatibility robustness with Eq.~\eqref{IR_SRc_NRc} by considering the example in Ref.~\cite{Cavalcanti16}. That is, Alice and Bob share a pure partially entangled state
\begin{equation}
|\phi\rangle = \cos\theta|00\rangle+\sin\theta|11\rangle\quad  \theta\in(0,\pi/4].
\label{Eq_pure_entangled_state}
\end{equation}
For this state, optimal measurements for Alice and Bob giving the maximal violation of the Bell-Clauser-Horne (CH) inequality~\cite{Clauser74} are known analytically (see, e.g., Ref.~\cite{Liang:PhDthesis}). One can, then, estimate the DI lower bounds on the incompatibility robustness of Alice's and Bob's measurements by using the above different approaches. The results are plotted in Fig.~\ref{Fig:IR}, together with our improved bound $\SRDIc$, that will be introduced below. With some attention, one observes a small but noticeable gap (of the order of $10^{-3}$ or less) between $\SRDI$ and $\NRc_\ell$ for some value of $\theta$, even though we already employed the $5$th level of AMM in our computation of $\SRDI$ (while the computation of $\NR^c$ was achieved using the $2$nd level of the NPA hierarchy).

Such a gap may be explained by the fact that  $\SRDI$ does not take into account of the consistency condition $\sum_a \tau_{a|x}=\sum_a \rho_{a|x}$, present in some form in $\NRc$, which provides a better lower bound to ${\rm IR}$. To improve our bound, we apply the the AMMs approach to $\SRc$. Then, the optimization problem of Eq.~\eqref{Eq_SR_c} gets relaxed to
\begin{equation}
\begin{aligned}
\min_{\{u_v\}} ~~&\left(\sum_{\lambda}\chi_{\mbox{\tiny DI}}^{(\ell)}[\sigma_{\lambda}]_\text{tr}\right)-1\\
\text{s.t.}~~ &\sum_{\lambda}D(a|x,{\lambda})\chi_{\mbox{\tiny DI}}^{(\ell)}[\sigma_{\lambda}] \succeq 
\chiDI \quad\forall~a,x,\\
&\sum_\lambda\chi_{\mbox{\tiny DI}}^{(\ell)}[\sigma_{\lambda}] = \sum_\lambda\chi_{\mbox{\tiny DI}}^{(\ell)}[\sigma_{\lambda}]_{\text{tr}} \cdot \sum_a \chiDI \quad\forall x,\\
&\chi_{\mbox{\tiny DI}}^{(\ell)}[\sigma_{\lambda}]\succeq 0\quad \forall ~\lambda\quad \forall ~\lambda,\\
&\sum_a\chiDI = \sum_a\chi_{\mbox{\tiny DI}}^{(\ell)}[\rho_{a|x'}] ~~\forall\, x\neq x',\\
&\chiDI\succeq 0 ~~\forall ~a,x,\\
&P(a,b|x,y)=P_{\mbox{\tiny obs}}(a,b|x,y)\quad\forall\quad a,b,x,y.
\end{aligned}
\label{Eq_SRDIc}
\end{equation}
This optimization problem, however, is not in the form of an SDP since the third line contains quadratic constraints in the free variables. To circumvent this complication, we can relax the original problem by keeping, instead, only a subset of the original constraints, i.e., entries
\begin{equation}
	\sum_\lambda\Big[\chi_{\mbox{\tiny DI}}^{(\ell)}[\sigma_{\lambda}]\Big]_{ij} = \sum_\lambda\chi_{\mbox{\tiny DI}}^{(\ell)}[\sigma_{\lambda}]_{\text{tr}}
	\cdot \sum_a \Big[\chiDI\Big]_{ij} \quad\forall\,\, x,
\label{Eq_SRDIc_constr}
\end{equation}
where $i,j$ are those corresponding to $[\chiDI]_{ij}=P(a,b|x,y)$. With this replacement, Eq.~\eqref{Eq_SRDIc} becomes an SDP, and we refer to its solution as $\SRDIcPobs$. Clearly, $\SRDIcPobs$ is a lower bound on $\SRcrax$ as it is obtained by solving a relaxation to the optimization problem of Eq.~\eqref{Eq_SRDIc}, and hence of Eq.~\eqref{Eq_SR_c}. At the same time, for any given level $\ell$, a straightforward comparison shows that the lower bound $\SRDIcPobs$ obtained by solving Eq.~\eqref{Eq_SRDIc} (with the third line replaced in the manner mentioned above) provides an upper bound on $\SRDIPobs$ obtained by solving Eq.~\eqref{Eq_relax_SR}, thus giving:

\begin{equation}
\IREax \geq \SRcrax \geq \SRDIcPobs \ge \SRDIPobs.
\label{LB_IR_AMM2}
\end{equation}
Table~\ref{TB_DIIRIW} summarizes the various approaches discussed above for the DI quantification of measurement incompatibility. From Fig.~\ref{Fig:IR}, we can see that $\SRDIc$ provides a much better bound (in some instances, even tight bounds) on $\IR$ compared to  $\SRDI$ and $\NRc_\ell$. {\blu On the other hand, it is also clear from the plots that, in these instances, $\SRDI$ already provides a tight bound on the underlying SR.}

\begin{center}
\begin{table}[h!]
\centering
\caption{Different methods that can be used to provide a DI quantification of measurement incompatibility.
}
\begin{tabular}{|c|c|}
\hline
method & bound relations  \\ \hline \hline
CS
& $ \IREax \geq \SRcrax \geq \NR^\text{c}_\ell(\Pobs) $ \\ \hline
CBLC
& $ \IREax \geq \SRrax \geq \SRDIPobs $\\ \hline
modified CBLC$^\dag$
& $ \IREax \geq \SRcrax \geq \SRDIcPobs $\\ \hline
\end{tabular}\label{TB_DIIRIW}
\end{table}
\end{center}

\section{Multipartite generalization and post-quantum steering}

%
%

Evidently, the framework of AMM introduced in Sec.~\ref{Sec:AMM} can be generalized to a scenario with more than two parties. Below, we discuss this specifically for the tripartite scenario and explain how this leads to novel insights on the set of correlations characterized by the framework of AMM.
%

\subsection{Steering in the tripartite scenario}
Following Ref.~\cite{Sainz15}, let us consider a tripartite Bell-type experiment where only Charlie has access to trusted (i.e., well-characterized) measurement devices. If we denote the shared quantum state by $\rabc$, the local POVM acting on Charlie's subsystem as $E^\tC_{c|z}$, then the analog of Eq.~\eqref{Eq:Quantum} reads as:
\begin{equation}\label{Eq:Quantum3}
	P(a,b,c|x,y,z)\stackrel{\Q}{=}\tr\left(\rabc\,E^\tA_{a|x}\otimes E^\tB_{b|y}\otimes E^\tC_{c|z}\right), 
\end{equation}
while that of Eq.~\eqref{Eq_quantum_assemblage} reads as:
\begin{equation}\label{Eq_quantum_assemblage3}
	\rho^\ttC_{ab|xy} = \tr_\text{A,B}(E_{a|x}^\tA\otimes E^\tB_{b|y}\otimes\id~\rabc)\quad \forall\,\, a,x, b,y.
\end{equation}
It is straightforward to see from Eq.~\eqref{Eq_quantum_assemblage3} that the assemblage $\{\rabxy\}_{a,b,x,y}$ (hereafter abbreviated as $\{\rabxy\}$) satisfy the positivity constraints and some no-signaling-like consistency constraints, i.e., 
\begin{equation}\label{Eq_valid_assemblage3}
\begin{aligned}
	 &\rabxy\succeq 0 \quad \forall\,\, a,b,x,y,\quad \tr\sum_{a,b}\rabxy =1,\\
	&\sum_{a}\rabxy = \sum_{a}\rho_{ab|x'y}^\ttC \quad\forall\, x, x', y,\\
	&\sum_{b}\rabxy = \sum_{b}\rho_{ab|xy'}^\ttC \quad\forall\, x, y, y',\\
	&\sum_{a,b}\rabxy = \sum_{a,b}\rho_{ab|x'y'}^\ttC \quad\forall\, x, x', y, y'.
\end{aligned}
\end{equation}
As with the bipartite case, the assemblage $\{\rabxy\}$ is said to admit an LHS model from A and B to C if there exists a collection of normalized quantum states {\blu $\{\hat{\sigma}_\lambda\}$}, probability distribution $P(\lambda)$, response functions $P(a|x,\lambda)$ and $P(b|y,\lambda)$ such that {\blu ${\rabxy = \sum_\lambda P(a|x,\lambda)P(b|y,\lambda)P(\lambda) \hat{\sigma}_\lambda}$ } for all $a,b,x,y$. Otherwise, the assemblage is said to be steerable from A and B to C.

\subsection{AMMs in a tripartite scenario}
\label{Sec:AMM3}
To generalize the AMM framework to the aforementioned steering scenario, consider the analog of Eq.~\eqref{Eq:LocalMapping}
that acts on the Hilbert space of Charlie's  system $\rho_\ttC$:
\begin{equation}\label{Eq:CPmap:C}
	\Lambda_\ttC(\rc) = \sum_n K_n\rc K_n^\dag,\quad
	K_n = \sum_i |i\rangle_{\bar{\text{C}}\text{C}} \langle n|C_i, 
\end{equation}

where $\{|i\rangle\}$ ($\{|n\rangle\}$) are orthonormal bases vectors for the output (input) Hilbert space $\bar{\text{C}}$ (C) and $C_i$, $C_j$ are some operators acting on C. 

Specifically, for each combination of outcome $a,b$ and setting $x,y$, applying the local CP map of Eq.~\eqref{Eq:CPmap:C} to the conditional state $\rabxy$ gives rise to a matrix of expectation values:
\begin{equation}\label{Eq_AMM3}
\begin{aligned}
	\chi[\rabxy,\{C_i\}] &= \Lambda_\ttC(\rabxy)\\
	&= \sum_{i,j}\ket{i}\!\bra{j}\tr[\rabxy C_j^\dagger C_i]\quad \forall\ a,b,x,y,
	\end{aligned}
\end{equation}
where $\{C_i\}$ are again operators formed from the product of $\{\id\}\cup\{E^\tC_{c|z}\}_{z,c}$. When $\{C_i\}$ involves operators that are at most $\ell$-fold product of Charlie's POVM elements, we say that the collection of matrices in Eq.~\eqref{Eq_AMM3} defines AMMs of level $\ell$, which we denote  by $\chil[\rabxy]$.


In a DI scenario, neither the assemblage $\{\rabxy\}$ nor the measurement assemblage $\{E_{c|z}^\ttC\}$ is assumed. Therefore, the level $\ell$ AMMs corresponding to $\chi[\rho_{ab|xy},\{C_i\}]$ in the DI setting, which we denote by $\chiDIabc$, is not fully determined. Following analogous procedure as that detailed in Sec.~\ref{Sec_AMMs_formulation}, one finds that the elements of the $\chiDIabc$ fall under two categories: observable correlation (i.e., conditional probabilities) $\Pobsabcxyz$\footnote{To save the space, $P(a,b,c|x,y,z)$ is abbreviated as $P(abc|xyz)$ when there is no risk of confusion.} and unknown variables. 

As an example, consider the steering scenario with binary input and output on Charlie's side such that $C_i\in\{\id,E_{1|1}^\tC,E_{1|2}^\tC\}$. Then, for all $a,b,x,y$, the first-level AMMs take the form of
\begin{equation}
\begin{split}
&\chiDIabcfirst =\\
&\begin{pmatrix}
\tr( \rabxy) & \tr( \rabxy E_{1|1}^\ttC) &  \rabxy E_{1|2}^\ttC)\\
\tr( \rabxy E_{1|1}^\ttC) & \tr \rabxy E_{1|1}^\ttC) & \tr(\rabxy E_{1|2}^{C\dag} E_{1|1}^\ttC)\\
\tr( \rabxy E_{1|2}^\ttC) & \tr( \rabxy E_{1|1}^{C\dag} E_{1|2}^\ttC)  & \tr( \rabxy E_{1|2}^\ttC)
\end{pmatrix}\\
&=\begin{pmatrix}
P_{\mbox{\tiny obs}}(ab|xy) & P_{\mbox{\tiny obs}}(ab1|xy1) & P_{\mbox{\tiny obs}}(ab1|xy2)\\
P_{\mbox{\tiny obs}}(ab1|xy1) & P_{\mbox{\tiny obs}}(ab1|xy1) & u_1^{abxy}\\
P_{\mbox{\tiny obs}}(ab1|xy2) & u_1^{abxy}  & P_{\mbox{\tiny obs}}(ab1|xy2)
\end{pmatrix},
\end{split}
\end{equation}
where we have made use of the simplification mentioned in Sec.~\ref{Sec:AMM} and expressed the experimentally inaccessible expectation value as:
\begin{equation}
\tr(\rabxy E_{1|2}^{\text{C}\dag} E_{1|1}^\text{C}) = \tr(\rho_{a|x}E_{1|1}^{\text{C}\dag} E_{1|2}^\text{C}) =u_1^{abxy},
\end{equation}
with $u_v^{abxy}\in\mathbb{R}$.

\subsection{Correlations characterized by the AMM framework and post-quantum steering}

In Ref.~\cite{SLChen16}, it was left as an open problem whether the set of correlations characterized by the AMM framework converges to the set of quantum distributions, i.e., the set of $\vecP$ that satisfy Born's rule. 
In this section, we show that in the tripartite scenario, the set of $\vecP$ allowed by demanding the positivity of AMMs---even in the limit of $\ell\to\infty$---generally cannot lead to the set of $\vecP$ that can be written in the form of Eq.~\eqref{Eq:Quantum3}.

To this end, we recall from Ref.~\cite{Sainz17} that there exists assemblage $\{\rabxy\}$ satisfying Eq.~\eqref{Eq_valid_assemblage3} but not Eq.~\eqref{Eq_quantum_assemblage3} for any $\rabc$ and any local POVM $\{E^\tA_{a|x}\}$, $\{E^\tB_{b|y}\}$. The authors of Ref.~\cite{Sainz17} dubbed this phenomenon {\em post-quantum steering}. A simple example of this kind is given by $\rabxy = \frac{1}{4}[1-(-1)^{ab+(x-1)(y-1)}]\hat{\rho}$ where $x,y\in\{1,2\}$, $a,b\in\{0,1\}$ and $\hat{\rho}$ is an arbitrary, but normalized density operator. Since the resulting marginal distribution $P(a,b|x,y)$ is exactly that of a Popescu-Rohrlich box~\cite{Popescu1994}, we see that this assemblage cannot have a quantum realization.

Now, note from our discussion in Sec.~\ref{Sec:AMM3} that if we start from an assemblage satisfying Eq.~\eqref{Eq_valid_assemblage3}, the resulting AMMs are always positive semidefinite, and hence are compatible with the physical requirements imposed on AMMs. However, as mentioned above, there exists assemblage $\{\rabxy\}$ satisfying Eq.~\eqref{Eq_valid_assemblage3} but which is not quantum realizable. We thus see that the AMM framework in the tripartite scenario, as described in Sec.~\ref{Sec:AMM3}, can, at best, lead to a characterization of the set of post-quantum-steerable correlations, i.e., a {\em superset} of correlations satisfying Eq.~\eqref{Eq:Quantum3} that also include, e.g., non-signaling, but stronger-than-quantum marginal distributions between A and B.

On the other hand, it follows from the results of Refs.~\cite{Gisin89,Hughston93} that the phenomenon of post-quantum steering cannot occur in the bipartite scenario. Thus the problem of whether the set of correlations characterized by the AMM framework leads to the set of quantum distributions remains open in the bipartite scenario. Likewise, if one {\blu considers} AMMs in a tripartite scenario based on one party steering the remaining two parties, the above argument {\blu does not apply either}. As such, the problem whether one recovers---in the asymptotic limit---the quantum set, cf. Eq.~\eqref{Eq:Quantum3}, using the AMM framework remains open.

\section{Concluding Remarks}
\label{Sec:Conclusion}

In this work, we have further explored and developed the AMM framework introduced in Ref.~\cite{SLChen16}. To begin with, we flashed out the details on how a DI bound on steering robustness (SR) provided by the AMM framework allows us to estimate the usefulness of an entangled state in the kind of subchannel discrimination problem discussed in Ref.~\cite{Piani15}. 

We then went on to compare the DI bound on the generalized robustness of entanglement provided by the AMM framework against that given by the approach of Cavalcanti and Skrzypczyk~\cite{Cavalcanti16}. Within our computational limit, the bounds of AMM appear to be {\blu slightly tighter than (or at least as good) } those from the latter approach. In the process, we also offered another mean to bound the generalized robustness of entanglement from the data alone via the approach of Moroder {\em et al.}~\cite{Moroder13}. This last set of DI bounds turned out to be much stronger than that offered by the other two approaches. In these comparisons, we considered the two-qudit isotropic states where we also evaluated their generalized robustness of entanglement explicitly (see Appendix~\ref{App:ERStates}).

Next, we compared the DI bound on the incompatibility robustness (IR) given by the AMM framework against that of Ref.~\cite{Cavalcanti16}. In this case, the DI bounds offered by the AMM approach---based on bounding SR---do not perform as well compared with those of Ref.~\cite{Cavalcanti16}, which are based on bounding the underlying consistent steering robustness. Motivated by this difference, we then provided an alternative way to lower bound---in a DI manner---the consistent steering robustness via the AMM framework. This turned out to provide---as compared with the approaches just mentioned---much tighter (and in some instances even tight) DI bounds on the underlying IR. Even then, let us note that, in general, a tight DI bound on the underlying IR does guarantee the possibility to self-test the underlying measurements, as exemplified by the results of Ref.~\cite{Andersson2017}.  On a related note, we demonstrated in Appendix~\ref{App:SW} how the AMM framework can be used to provide a DI lower bound on the steerable weight, and hence the incompatibility weight---another measure of incompatibility between different measurements.

We also briefly explored the framework in the tripartite scenario. This led to the observation that the AMM framework generally does not characterize the set of quantum correlations, but rather the set of correlations where the phenomenon of post-quantum steering is allowed. However, the problem of whether the set of correlations characterized by the AMM framework converges to the quantum set in the bipartite scenario, or in a multipartite scenario where one party tries to steer the states of the remaining parties remains unsolved.

\begin{acknowledgements}
We are grateful to Daniel Cavalcanti and Paul Skrzypczyk for useful discussions and for sharing their computational results in relation to the plot shown in Fig.~\ref{Fig:IR}. This work is supported by the Ministry of Science and Technology, Taiwan (Grants No. 103-2112-M-006-017-MY4, 104-2112-M-006-021-MY3, 107-2112-M-006-005-MY2, and 107-2917-I-564 -007 (Postdoctoral Research Abroad Program)), and by the FWF Project M~2107~(Meitner-Programm).

\end{acknowledgements}

\appendix

\section{Comparison of different moment-matrix approaches}\label{Sec_App_MMs}
The following table provides a comparison of different moment-matrix approaches.

\onecolumngrid
\vspace{\columnsep}
\begin{center}
\begin{table}[h!]
\centering
\caption{Different moment-matrix hierarchies in the context of (partially) device-independent (DI) characterizations. 
The first column labels the name of the author (or their initials if there is more than one author). The second column shows how the $\ell$-th level construction of the $\ell$-th level of each moment matrix (for simplicity, we provide an exemplification of the construction assuming two, or otherwise the minimal of parties where the hierarchy is applicable). Symbols like $E_{a|x}^\tA\in\mathsf{L}(\mathcal{H}_A)$ denotes the $a$-th POVM element of A's $x$-th measurement, while a POVM element like $E_{a|x}^\text{\tiny AB}\in\mathsf{L}(\mathcal{H}_{AB})$ acts on the global Hilbert space of A and B. For brevity, $b_i$ ($y_i$) etc. is introduced to denote one of the measurement outcomes (settings). $M_x^k$ denotes the $x$-th observable of party-$k$. }
\begin{tabular}{|c|c|c|l|}
\hline
Hierarchies & Moment matrix construction & Accessible data & Examples of applications \\ \hline \hline

NPA & $\Gamma_{ij}^{(\ell)}=\tr(\rab O_j^{(\ell)\dag} O_i^{(\ell)})$ & Correlation $\vecP$ & $\bullet$ Characterization of $\Q$ \\  \cite{NPA}
    & $\mathcal{O}^{(\ell)}=\id\cup\mathcal{S}^{(1)}\cup\mathcal{S}^{(2)}\cup ...\cup \mathcal{S}^{(\ell)}$ & (DI) & $\bullet$ Various DI characterizations \\ 
    & $\mathcal{S}^{(\ell)}=\{E_{a_1|x_1}^{AB}...E_{a_k|x_k}^{AB} E_{b_{k+1}|y_{k+1}}^{AB}...$& & \\
    & $...E_{b_\ell|y_\ell}^{AB}\}_{x=1...n_x~ a=1,...,n_a-1}^{y=1...n_y~ b=1,...,n_b-1}$, $0\leq k\leq \ell$ & & \\ \hline

MBLHG & $[\chi^{(\ell)}_\text{\tiny DI}]_{ijkl}=\tr(\rab A_k^{(\ell)\dag} A_i^{(\ell)} \otimes B_l^{(\ell)\dag} B_j^{(\ell)})$ & Correlation $\vecP$ & $\bullet$ Characterization of $\Q$ \\ \cite{Moroder13}
      & $\mathcal{A}^{(\ell)} = \id\cup\mathcal{S}_A^{(1)}\cup\mathcal{S}^{(2)}\cup ...\cup \mathcal{S}_A^{(\ell)}$ & (DI) & $\bullet$ DI lower bound on negativity, Hilbert space dim. \\
      & $\mathcal{B}^{(\ell)} = \id\cup\mathcal{S}_B^{(1)}\cup\mathcal{S}^{(2)}\cup ...\cup \mathcal{S}_B^{(\ell)}$ & & $\bullet$ Tsirelson bounds for PPT quantum states\\
      & $\mathcal{S}_A^{(\ell)}=\{E_{a_1|x_1}^A...E_{a_\ell|x_\ell}^A\}_{x=1,...,n_x}^{a=1,...,n_a-1}$ &  &  $\bullet$ DI lower bound on entanglement depth~\cite{Liang15} etc. \\
      & $\mathcal{S}_B^{(\ell)}=\{E_{b_1|y_1}^B...E_{b_\ell|y_\ell}^B\}_{y=1,...,n_y}^{b=1,...,n_b-1}$ &  &  $\bullet$ DI lower bound on ER (present work) \\ \hline

Pusey & $[\chi^{(\ell)}(\rab)]_{ij} = \tr_\tA(\rab O_j^{(\ell)\dag} O_i^{(\ell)}\otimes\id)$ & Assemblage  & $\bullet$ Characterization of quantum assemblages \\ \cite{Pusey13}
      & $\mathcal{O}^{(\ell)} = \id\cup\mathcal{S}^{(1)}\cup\mathcal{S}^{(2)}\cup ...\cup \mathcal{S}^{(\ell)}$ & $\{\rho_{a|x}\}$ & $\bullet$ 1-sided DI lower bound on negativity \\
      & $\mathcal{S}^{(\ell)}=\{E_{a_1|x_1}^A...E_{a_\ell|x_\ell}^A\}_{x=1,...,n_x}^{a=1,...,n_a-1}$ & (1-sided DI) & $\bullet$ Steering bounds for PPT quantum states\\ \hline

KSCAA & $\Gamma_{ij}^{(\ell)} = \tr(\rab O_i^{(\ell)\dag} O_j^{(\ell)})$ & Correlation $\vecP$ & $\bullet$ Characterization of unsteerable moments \\ \cite{Kogias15}
      & $\mathcal{O}^{(\ell)}=\id\cup\mathcal{S}^{(1)}\cup\mathcal{S}^{(2)}\cup ...\cup \mathcal{S}^{(\ell)}$  & (some $M^k_{x_k}$ & \\
      & $\mathcal{S}^{(\ell)}\!=\!\{ M_{x_1}^A...M_{x_\ell}^A\otimes\id$, $M_{x_1}^A...M_{x_{\ell-1}}^A\otimes M_{y_1}^B$, &  assumed ) & \\ 
      &  ..., $\id\otimes M_{y_1}^B...M_{y_\ell}^B \}_{x=1,...,n_x}^{y=1,...,n_y}$ & (partially DI)& \\ \hline

SBCSV & $\Gamma_{ijkl}^{(\ell)} = \tr(\rabc B_k^{(\ell)\dag} B_i^{(\ell)} C_l^{(\ell)\dag} C_j^{(\ell)})$ & Assemblage & $\bullet$ Characterization of quantum assemblages \\ \cite{Sainz15}
      & $\mathcal{B}^{(\ell)}=\id\cup \{E_{b_1|y_1}^{ABC}...E_{b_\ell|y_\ell}^{ABC}\}_{y=1,...,n_y}^{b=1,...,n_b-1}$ & $\{\rho_{bc|yz}\}$ &$\bullet$ Quantum bounds on steering inequalities \\
      & $\mathcal{C}^{(\ell)}=\id\cup\{E_{c_1|z_1}^{ABC}...E_{c_\ell|z_\ell}^{ABC}\}_{z=1,...,n_z}^{c=1,...,n_c-1}$ & ($n$-sided DI, $n\ge 2$) & \\ \hline

CBLC  & $[\chi_\text{\tiny DI}^\ell(\rho_a|x)]_{ji}=\tr(\rho_{a|x} O_i^{(\ell)\dag} O_j^{(\ell)})$ & Correlation $\vecP$  & $\bullet$ Outer approximation of $\Q$ \\ \cite{SLChen16}
      & $\mathcal{O}^{(\ell)} = \id\cup\mathcal{S}^{(1)}\cup\mathcal{S}^{(2)}\cup ...\cup \mathcal{S}^{(\ell)}$ & (DI) & $\bullet$ DI lower bound on steerability,  measurement \\ 
      & $\mathcal{S}^{(\ell)}=\{E_{b_1|y_1}^B...E_{b_\ell|y_\ell}^B\}_{y=1,...,n_y}^{b=1,...,n_b-1}$ & &  incompatibility, ER etc. \\ \hline

\end{tabular}
\end{table}\label{TB_MM_Refs}
\end{center}
\vspace{\columnsep}
\twocolumngrid

\section{Device-independent estimation of steerable weight}
\label{App:SW}

In this section we would like to show how the AMMs approach can be used to estimate the degree of steerability measured by the steerable weight~\cite{SNC14}. Given an assemblage, one can always represent it as a (possibly trivial) convex mixture of a steerable assemblage $\{\rho_{a|x}^{\text{S}}\}$ and an unsteerable assemblage $\{\rho_{a|x}^{\text{US}}\}$:
\begin{equation}\label{Eq_SW_def}
	\rho_{a|x} = (1-\mu) \rho_{a|x}^{\text{S}} + \mu \rho_{a|x}^{\text{US}}\quad \forall\,\, a,x.
\end{equation}
The steerable weight (SW) is the minimal weight associated with $\rho_{a|x}^{\text{S}}$ among all possible convex decompositions of $\{\rho_{a|x}\}$ according to Eq.~\eqref{Eq_SW_def}, i.e., $\text{SW} = \min (1-\mu)$. Importantly, SW that can be computed by solving the SDP~\cite{SNC14}:
\begin{subequations}
\begin{align}
{\rm SW}(\{\rho_{a|x}\}) = & \ \ \min_{\{\rho_\lambda\}} \ \  1 - \tr\sum_{\lambda}\rho_\lambda \label{Eq_SW1}\\
\text{s.t.}~
&\rho_{a|x} \succeq \sum_\lambda D(a|x,\lambda)\rho_\lambda  \quad \forall\  a,x.\label{Eq_SW2}\\
&\rho_\lambda \succeq 0 \quad \forall\ \lambda.\label{Eq_SW3}
\end{align}
\label{Eq_SW}
\end{subequations}

To obtain a lower bound on ${\rm SW}(\{\rho_{a|x}\})$ via the AMMs approach, we follow essentially the same steps in the derivation of Eq.~\eqref{Eq_relax_SR} to obtain the following SDP:
\begin{subequations}\label{Eq_relax_SW}
\begin{align}
\min_{\{u_v\}} ~~&1-\left(\sum_{\lambda}\chi_{\mbox{\tiny DI}}^{(\ell)}[\rho_{\lambda}]_\text{tr}\right) \label{Eq_relax_SW1}\\
\text{s.t.}~~ &\chiDI \succeq\sum_{\lambda}D(a|x,{\lambda})\chi_{\mbox{\tiny DI}}^{(\ell)}[\rho_{\lambda}]
 \quad\forall~a,x,\label{Eq_relax_SW2}\\
&\chi_{\mbox{\tiny DI}}^{(\ell)}[\rho_{\lambda}]\succeq 0\quad \forall ~\lambda,\label{Eq_relax_SW3}\\
&\sum_a\chiDI = \sum_a\chi_{\mbox{\tiny DI}}^{(\ell)}[\rho_{a|x'}] ~~\forall\, x\neq x',\label{Eq_relaxSW_nosig}\\
&\chiDI\succeq 0 ~~\forall ~a,x\label{Eq_relaxSW_posi},\\
&P(a,b|x,y)=P_{\mbox{\tiny obs}}(a,b|x,y)\quad\forall\quad a,b,x,y.
\end{align}
\end{subequations}
The lower bound on ${\rm SW}(\{\rho_{a|x}\})$ obtained by solving Eq.~\eqref{Eq_relax_SW} will be denoted by $\SWDIPobs$.

As with steering robustness, SW  provides~\cite{Cavalcanti16} a lower bound on a measure of the incompatibility between measurements, called ``incompatibility weight"~\cite{Pusey15} (IW). In analogy to Eq.~\eqref{Eq_SW_def}, IW is defined as the minimal weight associated with the non-jointly-mesurable assemblage in all possible convex decomposition of the given assemblage into a component that is jointly-measurable and one that is not. Since $\IWEax\geq\SWrax$,
$\SWDIPobs$ provides, via the analog of Eq.~\eqref{Eq:SRBounds} for SW, a DI lower bound on IW of the underlying measurement assemblage. In fact, as with the case of bounding IR by SR, our DI bounds on IW can be strengthened by introducing the additional constraints given by Eq.~\eqref{Eq_SRDIc_constr} in Eq.~\eqref{Eq_relax_SW}. We will denote the corresponding DI bounds by {\blu $\SWDIcPobs$.}

As an example, we may use the correlations $\vecP$ detailed in the caption of Fig.~\ref{Fig:IR} to estimate the IW of Bob's measurement assemblage as a function of $\theta$.  Our numerical results show that, whenever the state is entangled (i.e., $\theta\neq 0$), the DI bounds {\blu $\SWDIcPobs$ give } essentially the value 1, which is also the incompatibility weights of the underlying measurement assemblages.\footnote{The deviation from unity is negligible except when $\theta\approx 0$. But, as $\theta\to 0$, we may need to consider a higher-level of AMM relaxation to get a tight bound. For example, with $\ell=2$ and $\theta\gtrsim 1.5^\circ$, the deviation is within the numerical precision of the solver, but for $\theta\approx 0.75^\circ$, we see a deviation of the order of $10^{-4}$.  Alternative, tight DI lower bounds on IW for the same set of correlations can also be obtained via the consistent nonlocal weight introduced by Cavalcanti and Skrzypczyk~\cite{Cavalcanti16}.} 

\section{DI lower bounds on ER}

\subsection{Details of bounds for correlations obtained from qubit isotropic states}

\begin{figure}[h!]
\begin{center}
\emph{\includegraphics[width=9.5cm]{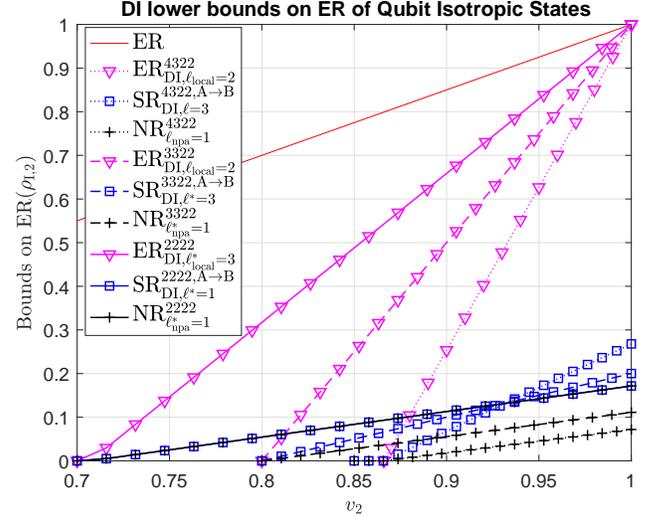} }
\caption{\label{Fig:FullDetails}
DI lower bounds on entanglement robustness ($\ER$) based on various Bell-inequality-violating correlations $\vecP$ obtained from the two-qubit isotropic states $\rho_{\text{\tiny I}, 2}(v_2)$  using the three approaches discussed in Sec.~\ref{Sec:DI-ER}. Specifically, these correlations  $\vecP$ are obtained from Eq.~\eqref{Eq:Quantum} for $\rho_{\text{\tiny I}, 2}(v_2)$ and measurements leading to their optimal CHSH-inequality violation (marked with ``2222" in the legend and solid line in the plot),  optimal $I_{3322}$-Bell-inequality violation  (marked with ``3322" in the legend and dashed line in the plot), and the optimal elegant-Bell-inequality-violation  (marked with ``4322" in the legend and dotted line in the plot) of these states. Bounds obtained from the approach of MBLHG~\cite{Moroder13}, AMM~\cite{SLChen16}, and of Ref.~\cite{Cavalcanti16} are marked, respectively, using  triangles ({\color{magenta} $\triangledown$}), squares ({\color{blue} $\square$}) , and crosses ($+$). For completeness, the actual value of $\ER[\rho_{\text{\tiny I}, 2}(v_2)]$ for each given value of visibility $v_2$, as given in  Eq.~\eqref{Eq:ER-rhoIso}, is also included as a (red) solid line.  $\ell$ in the legend denotes the level of the SDP hierarchy involved in the computation; a $*$ is included as a superscript of $\ell$ whenever the next level of the hierarchy, $\ell+1$, gives the same SDP bound (within a numerical precision of the order of $10^{-6}$).  
}
\end{center}
\end{figure}

\begin{figure}[h!]
\begin{center}
\emph{\includegraphics[width=9cm]{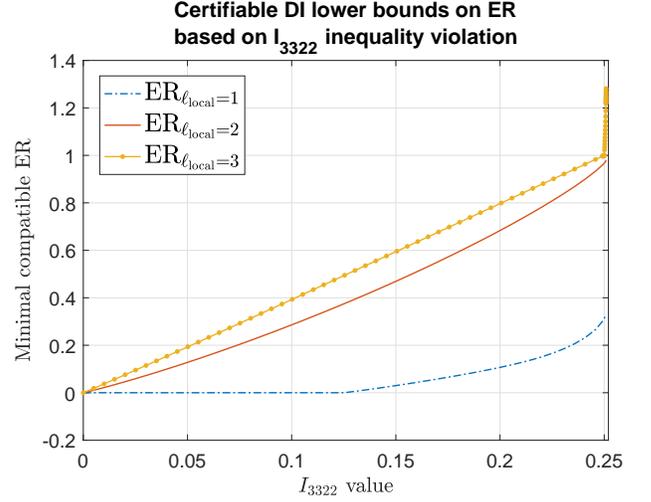} }
\caption{\label{Fig:i3322}
DI lower bounds on entanglement robustness ($\ER$) certified from the amount of $I_{3322}$ Bell-inequality violation based on a modification of the optimization given in Eq.~\eqref{Eq_DIER_MM} and a consideration of the hierarchy defined in Ref.~\cite{Moroder13}.}
\end{center}
\end{figure}

{\blu
\subsection{Bounds based on Bell-inequality violations}

Instead of solving Eq.~\eqref{Eq_DIER_MM}, we have also computed a relaxation thereof where we fixed only the value of specific Bell inequalities. For the case of the CHSH~\cite{Clauser69} Bell inequality, 
\begin{equation}
	\SCHSH:=\sum_{x,y=1,2} (-1)^{xy} E_{xy} \stackrel{\mathcal{L}}{\le } 2,
\end{equation}
where $E_{xy}:= \sum_{a,b=0,1} (-1)^{a+b} P(a,b|x,y)$, our numerical results suggest\footnote{Up to a numerical precision of 10$^{-7}$; likewise for the results obtained from considering the elegant Bell inequality given in Eq.~\eqref{Ineq:Elegant}.} the following {\em tight} lower bound:\footnote{To see that this and the following lower bounds are tight, it is sufficient to consider the two-qubit isotropic state, obtained by setting $d=2$ in Eq.~\eqref{Eq:IsoStates}.}
\begin{equation}
	\ER(\rho|\SCHSH=t)\ge \frac{t-2}{2\sqrt{2}-2},\quad 2\le t \le 2\sqrt{2}.
\end{equation}
On the other hand, for the elegant Bell inequality, 
\begin{equation}\label{Ineq:Elegant}
	\mathcal{S}_E:=\sum_{x=1}^4\sum_{y=1}^3 -(-1)^{\delta_{x,y+1}+\delta_{x,1}} E_{xy} \stackrel{\mathcal{L}}{\le } 6,
\end{equation}
where $\delta_{(,)}$ is the Kronecker delta function, we have instead the following {\em tight} lower bound on ER:
\begin{equation}
	\ER(\rho|\mathcal{S}_E=t)\ge \frac{t-6}{4\sqrt{3}-6},\quad 6 \le t \le 4\sqrt{3},
\end{equation}
where $\mathcal{S}_E$ is the observed value of the elegant Bell inequality violation.

For the $I_{3322}$ Bell inequality [see Eq.~(19) of Ref.~\cite{Collins04_0} for its explicit form], our numerical results up to level $\ell_{\rm local}=3$ are shown in Fig.~\ref{Fig:i3322}. With the highest-level relaxation that we have considered, the minimal value of $\ER$ compatible an $I_{3322}$ violation in the interval of $[0, 0.25]$ appears to be linear (up to a numerical precision of 10$^{-3}$); this lower bound can again be saturated by considering a two-qubit isotropic state in conjunction with its maximal quantum violation of the $I_{3322}$ inequality. However, we do not know if the nonlinear part of the curve where the Bell inequality is violated beyond $0.25$ can be saturated. In general, the fact that these DI lower bounds are saturated by the two-qubit isotropic state means that the results for ER$_\text{\tiny DI}$ shown in Fig.~\ref{Fig:FullDetails} can also be obtained by fixing the value of the observed Bell inequality, instead of considering the full set of probability distributions $\vecP$.}

\section{Generalized robustness of entanglement for the isotropic states}
\label{App:ERStates}

Here, we give a proof that the  generalized robustness of entanglement for the isotropic states ${\blu \rm{ER}}[\rId(v_d)]$ is indeed given by Eq.~\eqref{Eq:ER-rhoIso}. 
\begin{proof}
Suppose that for a given $\rId(v_d)$, the optimization problem of Eq.~\eqref{Eq_ER} is solved with  $(t^*,\tau^*_\text{\tiny AB})$ being an optimum solution. Since $\rId$ is invariant under an arbitrary local unitary transformation of the form $U\otimes \overline{U}$ (with $\overline{U}$ being the complex conjugate of $U$), we can see from Eq.~\eqref{Eq_ER} that, instead of $\tau^*_\text{\tiny AB}$, the local-unitarily-transformed state $U\otimes \overline{U}\tau^*_\text{\tiny AB} (U\otimes \overline{U})^\dag$ and $t^*$ must also form an optimum of the optimization problem. To see this, let $ {\blu \omega^*_\text{\tiny AB} }=\frac{\rId + t^*\tau^*_\text{\tiny AB}}{1+t^*}$, then $ {\blu \omega^*_\text{\tiny AB} }$ is separable by assumption, and thus $U\otimes \overline{U}{\blu \omega^*_\text{\tiny AB} } (U\otimes \overline{U})^\dag$ must also be separable for an arbitrary qudit unitary operator. Hence, instead of mixing $\rId$ with $\tau^*_\text{\tiny AB}$, we could just as well mix $\rId$ with $U\otimes \overline{U}\tau^*_\text{\tiny AB} (U\otimes \overline{U})^\dag$ in order to arrive at the minimum of Eq.~\eqref{Eq_ER}. More generally, given any optimum state $\tau^*_\text{\tiny AB}$ of the optimization problem, the twirled state $\int {\rm d}U\, U\otimes \overline{U}\tau^*_\text{\tiny AB} (U\otimes \overline{U})^\dag$ can also be used to arrive at the same optimum value $t^*$.

When performing the optimization of Eq.~\eqref{Eq_ER} with $\rab$ being an isotropic state, we can therefore, without loss of generality, restrict our attention to $\tab$s that are invariant under $U\otimes \overline{U}$-twirling, and hence by the characterization given in Ref.~\cite{Horodecki99} being an isotropic state. With this simplification, we may then rewrite Eq.~\eqref{Eq_ER} for the isotropic state as:
\begin{equation}\label{Eq_ER_rId}
\begin{aligned}
	\ER[\rIdv] = &\min_{t, u_d}  ~~t\geq 0\\
	\text{s.t.}& ~~\oab=\frac{\rIdv + t \rId(u_d)}{1+t}\quad \text{separable}.
\end{aligned}
\end{equation}

For an entangled isotropic state, i.e., one with $v_d>\frac{1}{d+1}$, it is easy to see---by invoking a convexity argument---that the minimum of the above optimization is attained by choosing $u_d$ such that $\rId(u_d)$ is separable and is furthest away from $\rIdv$ among all the separable $\rId(u_d)$. In other words, the optimization problem of Eq.~\eqref{Eq_ER_rId} is solved by setting $u_d=-\frac{1}{d^2-1}$. Equating the resulting mixture $\oab$ with an isotropic state that is barely separable, i.e., $\rId(\frac{1}{d+1})$ gives:
\begin{equation}
\begin{split}
	&\frac{v_d + tu_d}{1+t} = \frac{1}{d+1}\\
	\stackrel{u_d=-\frac{1}{d^2-1}}{\Rightarrow} & v_d - t\frac{1}{d^2-1}=\frac{1+t}{d+1}\\
	\Rightarrow &\, v_d - \frac{1}{d+1}=t\frac{d}{d^2-1}\\
	\Rightarrow &\, t=\frac{(d^2-1)v_d-(d-1)}{d}
\end{split}
\end{equation}
For a separable $\rIdv$, its generalized robustness of entanglement is easily seen to be $\ER[\rIdv]=0$. We thus arrive at the desired analytic expression of $\ER[\rIdv]=\max\left\{\frac{(d^2-1)v_d-(d-1)}{d},0\right\}$.


\end{proof}

\vskip 1cm


\bibliography{bib_AMM_Long}


\end{document}